\documentclass[aps,prb,twocolumn,groupedaddress]{revtex4-1}
\usepackage[bookmarks]{hyperref}
\usepackage[dvips]{graphicx}
\usepackage{amsmath}
\usepackage{amstext}
\usepackage{graphics}
\usepackage{exscale}
\usepackage{epsfig}
\usepackage{changebar}
\usepackage[T1]{fontenc}
\usepackage{bm}
\usepackage{bbm}

\usepackage{version}%\usepackage{psfrag}

\begin{document}
\title[]{Phase diagram and critical points of a double quantum dot} 
\author{Y Nishikawa${}^{1,2}$ D J G Crow${}^1$ and A C Hewson${}^1$ }
\affiliation{${}^1$Department of Mathematics, Imperial College, London SW7 2AZ,
  UK.}
\affiliation{${}^2$Graduate School of Science, Osaka City University, Osaka 558-8585, Japan} 
%\ead{a.hewson@imperial.ac.uk}
%ead{nisikawa@sci.osaka-cu.ac.jp}
%72.10.F Kondo transport
% 73.61 Quantum dots conductivity
% 71.10.A Fermi liquid theory
% 11.10.G Renormalization, field theory

\date{\today}

\begin{abstract}
We apply a combination of numerical renormalization group (NRG) and
renormalized
perturbation theory (RPT) to a model of two quantum dots (impurities)
described by two Anderson impurity models hybridized to their respective baths.
The dots are 
coupled via a direct interaction $U_{12}$ and an exchange interaction $J$.
The model has two types of quantum critical points, one at $J=J_c$ to a local
singlet state and one at $U_{12}=U_{12}^c$ to a locally charge ordered state.
The renormalized parameters which determine the low energy behavior are
calculated from the NRG.
The results confirm the  values predicted from the RPT on the approach to the critical
points, which can be expressed in terms of a single energy scale $T^*$ in all cases.
This includes cases without particle-hole symmetry, and cases with  asymmetry between the dots, 
where there is also a transition  at $J=J_c$. The results give a comprehensive
quantitative picture of the behavior of the model in the low energy Fermi
liquid regimes, and some of the conclusions regarding the
emergence of a single energy scale may apply to a more  general class of quantum
critical points, such as those observed in some heavy fermion systems.

\end{abstract}
\maketitle

\section{Introduction}
Experimental evidence in heavy fermion materials suggests that there is a close connection between
the regions of anomalous spin fluctuations 
associated with a quantum critical point and the occurrence of
superconductivity. There is a similar association in the cuprates \cite{Tai10} and iron pnictide \cite{Ste11}
superconductors. This has motivated many experimental studies of quantum behavior
in fermionic systems, which can be induced by lowering the transition in a
magnetically ordered state to zero by application of pressure, a magnetic
field
or by alloying\cite{SS10,LRVW07,CPSR01}. Many conjectures have been put forward to 
 understand this behavior, but as yet there is no fully quantitative or
comprehensive  explanation (for a recent review on this topic see reference\cite{SS10}). This situation
has led to a resurgence of interest in local or impurity models which
have quantum critical points. There are two reasons for this. Firstly,
impurity problems are simpler and well established techniques have been
developed for tackling them. Secondly, such systems can be simulated
using nanoscale devices, such as quantum dots, where the parameters
can be varied in a controlled way via gate voltages and the behavior examined
over different regimes. An  example is the two channel Kondo model,
which has a quantum critical point and non-Fermi liquid
behavior. An  arrangement of
 quantum dots and gate voltages has been used successfully  to test
 experimentally theoretical predictions for this model in 
the non-Fermi liquid regime\cite{PRSOD07}. \par
 The two impurity Kondo model is another example of a  model with
local  quantum critical behavior. Numerical renormalization group studies for
this model have
shown that there is a
 discontinuous 
transition on increasing the (antiferromagnetic) exchange interaction between the two impurities
from a state in which the impurity spins are predominantly screened by
the conduction electrons (Kondo screening) to one screened within a local
singlet state\cite{JV87}. There has been  recent experimental work\cite{BZD11} to examine the
critical behavior of this model using two magnetic impurities, one on a STM tip
and the other on a metal surface, such that the interaction between the two
impurities can be modified by controlling the distance of the STM tip from
the surface. Yet another type of critical point which has been studied theoretically\cite{GLK06}, 
one that could occur in two capacitatively
coupled quantum dots, has a discontinuous
transition to a  locally
charge ordered state. \par

In a recent paper we studied a two-impurity Anderson model
which has direct and exchange interactions between the two impurities
and displays both types of transition\cite{NCH12}. We showed, by using a combination
of renormalized perturbation theory (RPT) and numerical renormalization group
calculations (NRG), that we could derive many exact results for the low temperature
behavior in the Fermi liquid regime right up  to the quantum
critical points.  We were able to show the emergence of a single energy scale
on the approach to each type of quantum critical point with renormalized
Fermi liquid parameters corresponding
to a strong correlation regime. The results were restricted to a model with 
symmetry between the two impurities and particle-hole symmetry. In this paper
we enlarge upon those results, examine the phase diagram more fully,
and also relax the restrictions to include channel asymmetry and lack of
particle-hole symmetry.\par 

    \begin{figure}[!htbp]
   \begin{center}
     \includegraphics[width=0.4\textwidth]{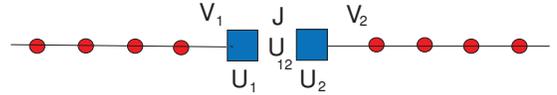}
     \caption{(Color online) A schematic illustration of  NRG chains for the
       double dot system. The quantum dots (filled squares)  interact via  an exchange $J$ and a direct term $U_{12}$, and are hybridized
with the first site of their respective conduction electron chains (filled circles). 
} 
     \label{chain3}
   \end{center}
 \end{figure}

 The model for the system we study  corresponds to two quantum dots connected via
 leads with their respective conduction baths. The Hamiltonian ${\cal H}$ for
 the system takes the form ${\cal H}={\cal H}_1+{\cal H}_2+{\cal H}_{12}$,
 where the  individual dots are
 described by a
single impurity Anderson model, so   ${\cal H}_\alpha$
for dot $\alpha=1,2$ is given by 
\begin{eqnarray}
&&{\cal H}_\alpha=\sum_{\sigma}\epsilon_{d,\alpha}d^{\dagger}_{\alpha,\sigma}d^{}_{\alpha,\sigma}+\sum_{k,\sigma}\epsilon_{k,\alpha}
c^{\dagger}_{k, \alpha,\sigma} c^{}_{k, \alpha,\sigma}\label{model1a} \\
&&+\sum_{k,\sigma} (V_{k,\alpha} d^{\dagger}_{\alpha,\sigma} c^{}_{k,\alpha,\sigma}
+  V_{k,\alpha}^* c^{\dagger}_{k, \alpha,\sigma} d^{}_{
  \alpha,\sigma})+U_\alpha n_{d,\alpha,\uparrow}n_{d,\alpha,\downarrow} \nonumber
\end{eqnarray}
where $d^{\dagger}_{ \alpha,\sigma}$, $d^{}_{ \alpha,\sigma}$, are creation and
annihilation operators for an electron at the impurity site in channel
$\alpha$, where $\alpha=1,2$,  and spin
component
$\sigma=\uparrow,\downarrow$.  
The creation and annihilation operators, $c^{\dagger}_{k,\alpha,\sigma}$,
$c^{}_{k,\alpha,\sigma}$, are for conduction electrons with energy
$\epsilon_{k,\alpha}$ in channel $\alpha$. The hybridization function for each
dot is given by $\Delta_{\alpha,\sigma}(\omega)=\pi\sum_k|V_{k,\alpha}|^2\delta(\omega-\epsilon_{k})$. We will make the
usual approximation and take  $\Delta_{\alpha,\sigma}(\omega)$  to be a constant
 $\Delta_\alpha$, independent of $\omega$ and $\sigma$, corresponding to the 
wide band limit with a constant density of states.
\par

The interaction between the two quantum dots is taken in the form of a direct
term $U_{12}$, and an exchange term $J$, so that  the interaction  Hamiltonian
${\cal H}_{12}$ has the form,
\begin{equation}
{\cal H}_{12}= U_{12}\sum_\sigma n_{d,1,\sigma} 
\sum_{\sigma'} n_{d,2,\sigma'}+
2J_{} {\bf S}_{d,1}\cdot{\bf S}_{d,2},  
\label{model2c}
\end{equation}
where $J>0$ for an antiferromagnetic coupling and  $J<0$ in the   ferromagnetic
case.\par 
Though the  model here is taken to describe a double quantum dot, it can serve
equally well to describe a number of other  physical situations. It can 
describe  a single magnetic impurity with two fold 'orbital'
degeneracy, where the exchange coupling $J$ then corresponds to a Hund's rule
term where a ferromagnetic exchange term would be appropriate
\cite{Yos76,NCH10s,NCH10a}.
In the strong coupling regime
 $U/\pi\Delta\gg 1$, the model corresponds to two Kondo models and with $U_{12}=0$ it has been much studied as a model of a system with two coupled
 magnetic impurities \cite{JV87,JV89,SVK93,Gan95}. 
The model with  $J=0$ has also been used to study  two
capacitatively coupled quantum dots \cite{GLK06}.\par
For the numerical renormalization group calculations (NRG) the conduction
electron states are transformed to a basis corresponding to a tight-binding
chain such that the model takes the form illustrated in Fig \ref{chain3}. The  
 densities of states
for the conduction electron baths (half-bandwidth $D$)  are discretized with a parameter $\Lambda
(>1)$ as in the original calculation of Wilson \cite{Wil75}
so that the couplings along the chain fall off as $\Lambda^{-N/2}$, for large
  $N$, where $N$ is the $N$th site along the chain from the dot. \par 

\section{Phase Diagram}

As the model used here is rather more general than those studied so far,
we show some results for the phase diagram which now includes
two QCPs. Before considering the general case we look at the case with
$U_{12}=0$. In the Kondo regime this is a much studied  model of a system with two coupled
 magnetic impurities \cite{JV87,JV89,SVK93,Gan95} which has a QCP at $J=J_c$
 where there is a
 sudden change in phase shift of the conduction electrons from $\pi/2$ to
 zero. When $J=0$ the two dots are isolated and their  
 ground states are (Kondo)  singlets giving a  phase shift of  $\pi/2$ in each
 channel. In the opposite limit
$J\to \infty$ the impurity spins form a decoupled {\em local} singlet such
that  the phase shift
of the 
conduction electrons in each channel
 is zero. In some related  models the change between
these two limits was found to occur continuously with increase of $J$.  Affleck, Ludwig
and Jones \cite{AL92,ALJ95}, however, cleared up the confusion resulting from
apparently conflicting results from the different models, in giving the conditions
for the change to occur discontinuously, resulting in   a QCP. 
 In a more recent study of  a quantum dot version of the  model, it 
has been shown that the transition occurs even when there is asymmetry  between the two
channels and away from particle-hole symmetry\cite{ZCSV06}.  For the calculations of the phase diagram we assume both channel
symmetry   $U_1, V_1=U_2,V_2$
and particle-hole symmetry $\epsilon_{d,\alpha}=-U_{\alpha}/2$, though  we relax some of these
conditions later. 
It will be useful to introduce a renormalised quantity $\tilde\Delta(0)$, which will be
defined
more fully later, such that
$\tilde\Delta(U,0)\to \Delta$ for $U\to 0$ and in
the Kondo regime $U/\pi\Delta\gg 1$,
 $\pi\tilde\Delta(U,0)\to 4T_{\rm K}$, where $T_{\rm K}$   the
Kondo temperature of the isolated dot, defined by the relation $T_{\rm K}=(g\mu_{\rm B})^2/4\chi_{\rm
  imp}$,  where $\chi_{\rm
  imp}$ is the $T=0$ spin susceptibility  for an isolated  dot.
\par

\par
  \begin{figure}[!htbp]
   \begin{center}
     \includegraphics[width=0.4\textwidth]{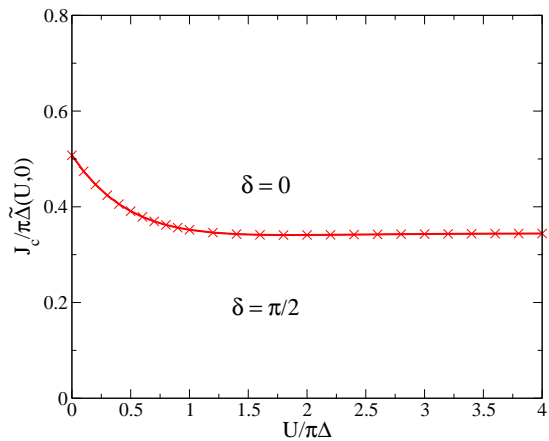}
     \caption{(Color online) A plot of  $J_c/\pi\tilde\Delta(U,0)$ as a
       function of  $ U/\pi\Delta$ for  $\pi\Delta=0.01$.
} 
     \label{Jc}
   \end{center}
 \end{figure}
 \begin{figure}[!htbp]
   \begin{center}
     \includegraphics[width=0.4\textwidth]{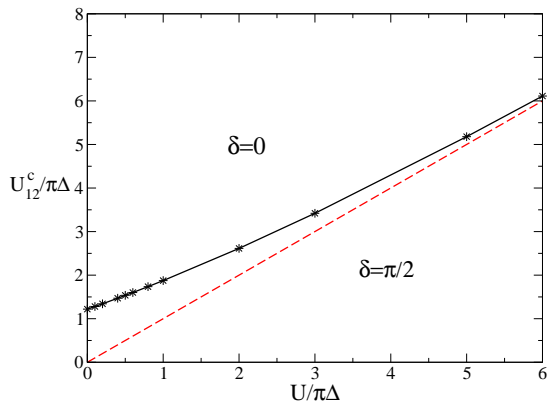}
     \caption{(Color online) A plot of  $U^c_{12}/\pi \Delta$, as a
       function of  $ U/\pi\Delta$ for the model with $J=0$.
} 
     \label{U12c}
   \end{center}
 \end{figure}

In Fig. \ref{Jc} we give a plot of the critical value $J_c/\pi\tilde\Delta(U,0)$ as a function
of $U/\pi\Delta$ for $\pi\Delta=0.01$ (note: we take this value for $\pi\Delta$ all subsequent
calculations unless specified explicitly). The transition in the NRG
calculations corresponds to a sudden change in the low energy fixed point from that of a
free chain with $N$ sites, where $N$ is even (odd) to 
one with $N$ odd (even). It can be seen that there is a transition for all
values of $U$ including $U=0$.
 For $U/\pi\Delta>3$ the curve is flat 
indicating that in this regime $J_c$ is proportional to $T_{\rm K}$,
corresponding to $J_c= 1.378T_{\rm K}$. 
 These results are  in line with calculations based
on the Kondo model where values of $J_c/T_{\rm K}\approx 1.4$ were estimated when expressed
using our definition of $T_{\rm K}$ \cite{JV87,JV89,SVK93} (in previous studies $T_{\rm K}$ was defined
by $\chi_s=2\mu_{\rm B}^2/\pi T_{\rm K}$ giving in our case  $J_c/T_{\rm
  K}=2.165$). \par

The second type of transition to a local charge ordered state occurs in the model  $J=0$ when 
$U_{12}$ reaches a critical value $U^c_{12}$. For $U_{12}=0$ the charges on
the two impurities are identical, but on increasing $U_{12}$ such that  $U_{12}\gg U$ it becomes
energetically
unfavourable and there is a local broken symmetry resulting in a charge
imbalance between the impurities. If there are matrix elements connecting
the two possible broken symmetry states to restore the symmetry then they must
be exceedingly small because the NRG results indicate a precise transition
at a critical value    $U_{12}=U^c_{12}$.
We plot the value of
$U^c_{12}/\pi\Delta$ as a function of  $U/\pi\Delta$ in Fig. \ref{U12c}. This
transition  exists for all values of $U$ and for $U/\pi\Delta\gg
3$ the critical values approach the value of $U$, i.e.  $U^c_{12}\to U$. 
 These results are in agreement with those given earlier by Galpin et al. \cite{GLK06}.\par

\begin{figure}[!htbp]
   \begin{center}
     \includegraphics[width=0.45\textwidth]{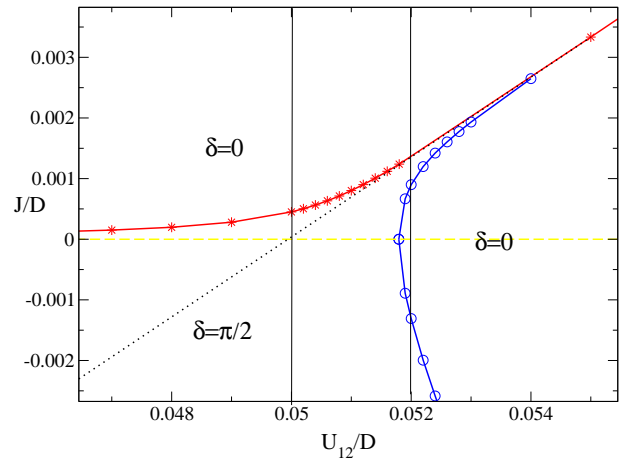}
     \caption{(Color online) A plot of the critical lines (local singlet-
       stars, local charge order - circles) separating
the regions with phase shift $\delta=\pi/2$ and $\delta=0$ as a function of
 $J/D$
and  $U_{12}/D$ for  $U/\pi\Delta=5$. The dotted line corresponds to
 $U_{12}=U+3J/2$. The two vertical lines relate to results shown later in
 Fig. \ref{u5_v5_vs_j} and Fig. \ref{u5_v5.2_vs_j}. }
     \label{phased2}
   \end{center}
 \end{figure}

The  phase diagram for the model with finite $J$ and $U_{12}$,
where both types of transitions occur, is shown in Fig.\ref{phased2}   
for the value $U/\pi\Delta=5$. The two types of transition are separated by
the line $U_{12}=U+3J/2$ along which the phase shift remains at the value
$\delta=\pi/2$. The two types of transition asymptotically approach this line on opposite
sides. \par
 To investigate the low energy behavior of the model in detail,
we first derive a number of exact results in terms of  renormalized parameters
using a renormalized perturbation theory (RPT) for the low energy properties in the Fermi liquid
regime.   
An analysis of the low energy  NRG fixed point is then used to calculate
 the renormalized
parameters in terms of the `bare' parameters that define the model. Once the
renormalized parameters have been calculated,  they can be substituted into
the equations for the exact evaluation of the  low energy thermodynamics.
The relevant  
equations based on the renormalized pertubation theory (RPT) will be  derived in the next section.\par

\bigskip
 \noindent

\section{RPT Analysis}

We start with the Fourier transform of the single particle Green's function
for the impurity $d$-state,
\begin{equation}
G_{d,\alpha,\sigma}(\omega_{n'})=-\int_0^\beta \langle T_\tau
d^{}_{\alpha,\sigma}(\tau)d^{\dagger}_{\alpha,\sigma}(0)\rangle e^{i\omega_{n'}\tau}\, d\tau,\label{gfdef}
\end{equation}
where $\omega_{n'}=(2n'+1)/\beta$ and $\beta=1/T$ and the brackets $\langle ...
\rangle$ denote a thermal average. 
This Green's function can be expressed in the form, 
\begin{equation}
G_{d,\alpha,\sigma}(\omega_n)={1\over i\omega_n-\epsilon_{d,\alpha} +i\Delta_{\alpha,\sigma}{\rm sgn}(\omega_n)-\Sigma_{\alpha,\sigma}(\omega_n)},\label{gf}
\end{equation}
where $\Sigma_{\alpha,\sigma}(\omega_n)$ is the self-energy.\par 
 For the zero
temperature Green's function, which will be our main concern, 
$\omega_n$ can be replaced by a  continuous variable $\omega$, and summations
over $\omega_n$ replaced by integrations over $\omega$.
For the renormalized perturbation theory \cite{Hew93}, the  Green's function in Eq.(\ref{gf}) can be re-expressed as $G_{d,\alpha,\sigma}(\omega_n)=z_\alpha\tilde G_{d,\alpha,\sigma}(\omega_n)$,
where $\tilde G_{d,\alpha,\sigma}(\omega_n)$ is the quasiparticle Green's function
given by 
\begin{equation}
\tilde G_{d,\alpha,\sigma}(\omega_n)={1\over i\omega_n-\tilde\epsilon_{d,\alpha} +i\tilde\Delta_\alpha{\rm sgn}(\omega_n)-\tilde\Sigma_{\alpha,\sigma}(\omega_n)}\label{qpgf}
\end{equation}
and the renormalized parameters, $\tilde\epsilon_{d,\alpha}$ and $\tilde\Delta_\alpha$
are given by
\begin{equation}\tilde\epsilon_{d,\alpha} =z_\alpha(\epsilon_{d,\alpha}+\Sigma_{\alpha,\sigma}(0)),\quad \tilde\Delta_\alpha
=z_\alpha\Delta_\alpha,\label{rself}
\end{equation}
where $z_{\alpha}=1/(1-\partial\Sigma_{\alpha,\sigma}(\omega)/\partial (i\omega))$
evaluated at $\omega=0$.
The quasiparticle self-energy $\tilde
\Sigma_{\alpha,\sigma}(\omega)$ is given by
\begin{eqnarray}&&\tilde\Sigma_{\alpha,\sigma}(\omega)
  =\nonumber\\&&z_\alpha\left(\Sigma_{\alpha,\sigma}(\omega)-\Sigma_{\alpha,\sigma}(0)\nonumber
-i\omega{\partial\Sigma_{\alpha,\sigma}(\omega)\over
\partial i\omega}\Big|_{\omega=0}\right ),
\end{eqnarray}
where we have assumed  the Luttinger theorem \cite{Lut60}, ${\rm Im}\Sigma(0)=0$, so that
${\rm Im}\tilde\Sigma_{\alpha,\sigma}(\omega)\sim \omega^2$ as $\omega\to 0$.
When expressed in this form, the $\omega=0$ part of the self-energy and its
derivative have been absorbed into renormalizing the parameters
 $\epsilon_{d,\alpha}$ and $\Delta_\alpha$,  so in setting up the perturbation 
expansion any further renormalization of these terms must be excluded, or
it will result in over-counting. In working with the fully renormalized
quasiparticles it is appropriate to use
the renormalized  or effective interactions between the quasiparticles.
In the single channel case, we identify the   renormalized interaction $\tilde U_\alpha$
as  the local four vertex
$\Gamma^{\alpha}_{\uparrow,\downarrow,\downarrow,\uparrow}(\omega_1,\omega_2,\omega_3,\omega_4)$ 
in the zero frequency limit \cite{Hew93},
\begin{equation}
\tilde
U_\alpha=z_\alpha^2\Gamma^{\alpha}_{\uparrow,\downarrow,\downarrow,\uparrow}(0,0,0,0).
\end{equation}
  In this case we need to include the 
local inter-channel 
four vertex, $\Gamma^{\alpha\alpha'}
(\omega_1,\omega_2)$
 corresponds to the  Fourier coefficient of the connected skeleton diagram for the two particle Green's
 function,
\begin{equation}
\langle T_\tau  {\bf S}_{d,1}(\tau_1)\cdot {\bf S}_{d,2}(\tau_2)\rangle,
\end{equation}
with the external legs removed.  We define
\begin{equation}
2\tilde
J=z_1 z_2\Gamma_{{\bf \sigma_1},{\bf \sigma}_2}(0,0,0,0).
\end{equation}
and we can define a renormalized direct interaction $\tilde U_{12}$ in
a similar way.

We can combine these terms to define a quasiparticle Hamiltonian $\tilde {\cal
  H}=\sum_\alpha\tilde{\cal H}_\alpha+\tilde{\cal H}_{12}$, where
\begin{eqnarray}
&&\tilde{\cal H}_\alpha=\sum_{\sigma}\tilde\epsilon_{d,\alpha} \tilde d^{\dagger}_{\alpha,\sigma}\tilde d^{}_{\alpha,\sigma}+\sum_{k\sigma}\epsilon_{k,\alpha}
c^{\dagger}_{k,\alpha,\sigma} c^{}_{k,\alpha,\sigma}\nonumber \label{qpmodel1a}\\
&&+\sum_{k,\sigma} (\tilde V_k \tilde d^{\dagger}_{\alpha,\sigma} c^{}_{k,\alpha,\sigma}
+ \tilde V_k^* c^{\dagger}_{k,\alpha,\sigma} \tilde d^{}_{ \alpha,\sigma})+
\tilde U_\alpha:\tilde n_{d,\alpha,\uparrow}\tilde n_{d,\alpha,\downarrow}:
\nonumber
\end{eqnarray}
and
\begin{equation}
\tilde{\cal H}_{12}= \tilde U_{12}:\sum_\sigma\tilde n_{d,1,\sigma} 
\sum_{\sigma'} \tilde n_{d,2,\sigma'}:+
2\tilde J_{} :\tilde{\bf S}_{d,1}\cdot\tilde {\bf S}_{d,2}:,  
\label{qpmodel2c}
\end{equation}
The brackets :${\hat O}$: indicate that the operator ${\hat O}$
within the brackets must be normal ordered with respect to the ground state of
the interacting system, which  plays the role of  the vacuum. 
This is because the
interaction terms only come into play when more than one quasiparticle is
created from the vacuum.

\par
The renormalized Hamiltonian is not equivalent to the original model, and
the relation between the original and renormalized model is best expressed
in the Lagrangian formulation, where frequency enters explicitly \cite{Hew01}. 
If the Lagrangian density ${\cal L}(\epsilon_d,\Delta_\alpha,U,J_{})$ describes the original model,
then by suitably re-arranging the terms we can write
\begin{equation}
{\cal L}(\epsilon_d,\Delta,U,J, U_{12})={\cal L}(\tilde\epsilon_{d},\tilde\Delta_\alpha,\tilde
U_\alpha,\tilde J,\tilde U_{12})+{\cal L}_c(\{\lambda_i\}),
\end{equation}
where the remainder part ${\cal L}_c(\{\lambda_i\})$
is known as the counter term. The set of coefficients  $\{\lambda_i\}$, 
 where $i=1,2,3,4,5$, can be expressed explicitly in terms
of the self-energy terms and vertices at zero frequency. These relations, however, are
not useful in carrying out the expansion. We want to work entirely with the
renormalized parameters and carry out the expansion in powers of
 $\tilde U_\alpha$, $\tilde J$ and $\tilde U_{12}$.  We assume that the $\lambda_i$
can be expressed in powers of quasiparticle interaction terms,  $\tilde U_\alpha$,
$\tilde J$ and $\tilde U_{12}$,
and can be determined order by order from the conditions that there should be no further renormalization of quantities which have already  been taken to be
fully renormalized.  These conditions are
\begin{equation}
\tilde\Sigma_{\alpha,\sigma}(0)=0,\quad {\partial\tilde\Sigma_{\alpha,\sigma}(\omega)\over
\partial i\omega}\Big|_0=0,
\end{equation}
and that the renormalized local and intersite 4-vertices at zero frequency are unchanged.

The propagator in the RPT is the free quasiparticle Green's function,
\begin{equation}
\tilde G^{(0)}_{d,\alpha,\sigma}(\omega_n)={1\over
  i\omega_n-\tilde\epsilon_{d, \alpha}
  +i\tilde\Delta_\alpha{\rm sgn}(\omega_n)}.\label{fqpgf}
\end{equation}
\par 
From Fermi liquid theory,  the quasiparticle interaction terms
 do not contribute to the
linear  specific heat coefficient $\gamma$ of the electrons. 
The specific heat coefficient $\gamma$ is given by
\begin{equation}
 \gamma=2\pi^2\sum_\alpha\tilde \rho_\alpha(0)/3,
\label{gam}
\end{equation} 
where $\tilde\rho_\alpha(\omega)$ is the free 
quasiparticle density of states per single spin and channel,
\begin{equation}
\tilde \rho_{\alpha}(\omega)={\tilde\Delta_\alpha/\pi\over (\omega-\tilde\epsilon_{d,\alpha})^2 +\tilde\Delta_\alpha^2}.\label{fqpdos}
\end{equation}
 The results for the spin
susceptibility  is given by 
\begin{equation}
\chi_s=2\mu_{\rm B}^2\sum_\alpha\tilde\eta_{s,\alpha}\tilde \rho_\alpha(0),\quad\chi_c=2\sum_\alpha\tilde\eta_{c,\alpha}\tilde \rho_\alpha(0),
\label{chis}
\end{equation}
where 
\begin{eqnarray}
\tilde\eta_{s,\alpha}=1&+&\tilde U_\alpha\tilde\rho_\alpha(0)-\tilde
J\tilde\rho_\beta(0), \nonumber \\
\tilde\eta_{c,\alpha}=1&-&\tilde U_\alpha\tilde\rho_\alpha(0) -2\tilde U_{12}\tilde\rho_\beta(0),
\label{etas}
\end{eqnarray}
where $\beta\ne\alpha$. The equations for the spin and charge susceptibilities, 
$\chi_s$ and $\chi_c$, can be proved using the Ward identities which follow
from
conservation of total spin and charge. They also correspond to a simple mean
field approximation to first order in the
effective interactions, $\tilde U_\alpha$, $\tilde J$ and
 $\tilde U_{12}$, 
  in the quasiparticle Hamiltonian 
Eqn. (\ref{qpmodel2c}). The corresponding equations for the staggered
susceptibilities also follow from these mean field equations and in the
Appendix we give an alternative way they can be derived from the leading
correction terms to the NRG fixed point,
\begin{equation}
\chi^{st}_{s}=2\mu_{\rm B}^2\sum_{\alpha}\tilde\eta^{st}_{s,\alpha}\tilde
\rho_\alpha(0),\quad\chi^{st}_{c}=2\sum_\alpha\tilde\eta^{st}_{c,\alpha}\tilde
\rho_\alpha(0),
\end{equation}
where
\begin{eqnarray}
\tilde\eta^{st}_{s,\alpha}=1&+&\tilde
U_\alpha\tilde\rho_\alpha(0)+\tilde J\tilde\rho_\beta(0),\nonumber \\
\tilde\eta^{st}_{c,\alpha}=1&-&\tilde U_\alpha\tilde\rho_\alpha(0) +2\tilde U_{12}\tilde\rho_\beta(0).
\label{chic-st}
\end{eqnarray}

From the equation of motion of  the Green's function
$G^{\alpha}_{k\sigma,k',\sigma}(\omega)$, which is the Fourier transform 
 of the retarded conduction electron
Green's function
$<<c_{\alpha,k,\sigma}^{}(t):c_{\alpha,k',\sigma}^{\dagger}(t')>>$ in channel $\alpha$, we can derive
the equation, 
\begin{equation}G^{\alpha}_{k\sigma,k',\sigma}(\omega)
={\delta_{k,k'}\over (\omega-\epsilon_k)}+{1\over
  (\omega-\epsilon_k)}t^{k,k'}_{\alpha,\sigma}(\omega)
{1\over (\omega-\epsilon_{k')}},
\label{eom}\end{equation}
where the t-matrix, $t^{k,k'}_{\alpha,\sigma}(\omega)$ is given by $t^{k,k'}_{\alpha,\sigma}(\omega)
  =V_{k,\alpha} G_{d,\alpha,\sigma}(\omega)V_{k',\alpha}$.
From the definition of the phase shift as the phase of the t-matrix for
  $\omega=0$,
\begin{equation}\delta_{\alpha,\sigma}={\rm tan}^{-1}\left({{\rm Im}\, G_{d, \alpha,\sigma}(0)\over {\rm Re}\,G_{d,\alpha,\sigma}(0)
   }\right),
\end{equation}
 we find
\begin{equation}
 \delta_\alpha-{\pi\over 2}=-{\rm tan}^{-1}\left
({\epsilon_{d,\alpha}+\Sigma_\alpha(0)\over \Delta_\alpha}\right)=-{\rm tan}^{-1}\left
({\tilde\epsilon_{d,\alpha}\over \tilde\Delta_\alpha}\right).
\label{qpocc}
\end{equation}
The total occupation of the impurity sites  $\sum n_{d,\alpha}=\sum\delta_\alpha/\pi$ at
$T=0$, which corresponds to the Friedel sum rule.\par

We consider first of all the model with $U_{12}=J=0$, which corresponds
to two independent Anderson models. With particle-hole symmetry
$\epsilon_{d,\alpha}=-U_\alpha/2$, $\tilde\epsilon_{d,\alpha}=0$, which corresponds to a phase shift
$\delta_\alpha=\pi/2$ in a given spin channel, independent of the value of $U_{\alpha}$.
 This leaves  just two renormalized parameters per dot, $\tilde \Delta_{\alpha}$
and $\tilde U_{\alpha}$, which are related in the strong coupling regime, as there is
only one  
relevant low energy scale, the Kondo temperature $T_{\rm
  K,\alpha}$. The relation takes the form, 
$\tilde U_\alpha=\pi\tilde\Delta_\alpha=4T_{\rm K,\alpha}$, where $T_{\rm K,\alpha}$ is
defined such that spin susceptibility of a single impurity is given by
$\chi_{s,\alpha}=(g\mu_{\rm B})^2/4T_{\rm K,\alpha}$. \par
On the approach to a quantum critical point we expect both
quasiparticle weight factors, $z_1\to
0$
and $z_2\to 0$,  as a singularity develops in the self-energy of the dot Green's
functions, which implies a divergence in the specific heat coefficient
$\gamma$. It also implies a divergence in the quasiparticle density of states
at the Fermi level, $\tilde\rho_\alpha(0)\to\infty$. As $J\to J_c$ we expect a divergence in the staggered spin susceptibility
 $\chi_s^{st}$ as it is the fluctuations in this channel that become critical.
We expect the other susceptibilities, 
$\chi_s$, $\chi_c$ and $\chi_c^{st}$, to remain finite, which leads us to
 require $\eta_{s,\alpha}\to 0$, $\eta_{c,\alpha}\to 0$ and $\eta_{c,\alpha}^{st}\to 0$ 
as  $J\to J_c$. These conditions give sufficient equations for us to  determine all the renormalized
parameters $\tilde\rho_\alpha(0)$, $\tilde U_\alpha$, $\tilde J$
and  $\tilde U_{12}$ in terms of a single energy scale. We find
\begin{equation} 
\tilde U_\alpha\tilde\rho_{\alpha}(0) \to 1,\quad \tilde
  J\tilde\rho_{\alpha}(0)
\to 2,\quad \tilde
  U_{12}\tilde\rho_{\alpha}(0)\to 0.\label{pred1}
\end{equation}
For the local charge transition it is only the  the staggered charge
susceptibility $\chi_c^{st}$ which should diverge as  $U_{12}\to U^c_{12}$.
 For the other susceptibilities to remain finite,  $\eta_{s,\alpha}\to 0$,
$\eta_{c,\alpha}\to 0$ and $\eta_{c,\alpha}^{st}\to 0$ 
as  $U_{12}\to
U^c_{12}$. This implies
\begin{equation} 
\tilde U_\alpha\tilde\rho_{\alpha}(0)\to -1,\quad \tilde
  J \tilde\rho_{\alpha}(0)
\to 0,\quad \tilde
  U_{12}\tilde\rho_{\alpha}(0)\to 1.\label{pred2}
\end{equation}
We can define an energy scale $T^*$ as $J\to J_c$ 
on the approach to the QCPs via  $1/\tilde\rho(0)=4T^*$ which for
particle-hole symmetry becomes   $\pi\tilde\Delta=4T^*$, such that the 
renormalized interaction parameters can be expressed in terms of  this single energy
scale $T^*$.\par\begin{figure}[!htbp]
   \begin{center}
     \includegraphics[width=0.4\textwidth]{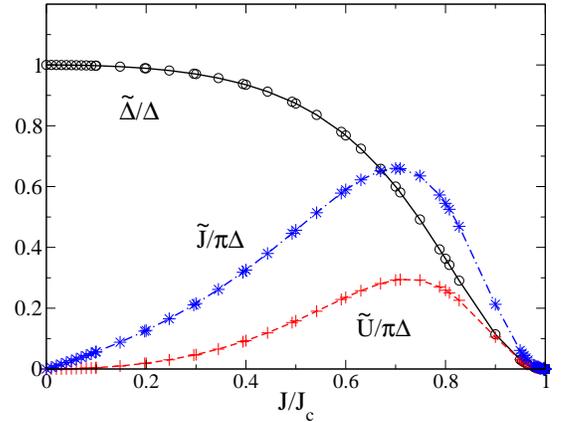}
     \caption{(Color online) A plot of $\tilde\Delta/\Delta$,
 $\tilde U/\pi\Delta$ 
and  $\tilde J/\pi\Delta$ 
as a function of $J/J_c$, for  $ U=0$.
}
   \label{rp_U_0.0}
   \end{center}
 \end{figure}

\bigskip
 \begin{figure}[!htbp]
   \begin{center}
     \includegraphics[width=0.4\textwidth]{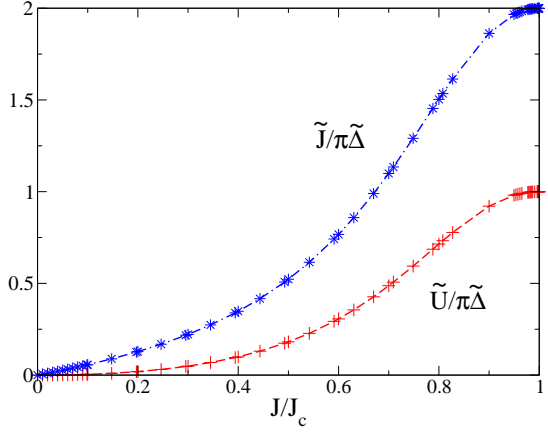}
     \caption{(Color online) A plot of the ratios,
 $\tilde U/\pi\tilde\Delta$ and,
and  $\tilde J/\pi\tilde\Delta$ 
as a function of $J/J_c$, for  $ U=0$.}
     \label{rp_ratios_u_0}
   \end{center}
 \end{figure}

Other exact results can be obtained from the RPT. An exact expression for the  renormalized self-energy for the model with particle-hole symmetry
$\tilde\Sigma(\omega)$ to order $\omega^2$ and $T^2$ follows from an RPT
calculation\cite{NCH12}  taken to second order in interaction parameters $\tilde U$,  $\tilde J$
and  $\tilde U_{12}$, and is given by 
\begin{equation}
\tilde\Sigma(\omega,T)={-i\pi^2I\tilde\Delta \over 64}\left[\left({\omega\over
      T^*}\right)^2+\left({\pi T\over T^*}\right)^2\right],\label{renself}
\end{equation}
where $I=(2\tilde U^2+3\tilde J^2+4\tilde U_{12}^2)/(\pi\tilde\Delta)^2$,
so $I\to 14$ as $J\to J_c$ and $I\to 6$ as $U\to U_{12}^c$,
and we have taken the channel symmetric case, $U_1=U_2$, $V_{1,k}=V_{2,k}$.
This is a generalization of the calculations given earlier in a
paper on an impurity model with a Hund's rule term \cite{NCH10s,Yos76}.  
 On substituting (\ref{renself}) into the quasiparticle
Green's function, we deduce  $\omega^2$ and $T^2$ contribution to the t-matrix,
$t_\alpha(\omega,T)$,
\begin{equation}
{i\pi D \over 16}\left[(I+4)\left({\omega\over
      T^*}\right)^2+I\left({\pi T\over T^*}\right)^2\right].
\end{equation}
The ratio of the $\omega^2$ to the $\pi^2T^2$ term is $(I+4)/I$,
which for $I=14$ gives $9/7$ in agreement with the calculation of 
Mitchell and Sela\cite{MS12} for the two impurity Kondo model. \par
To evaluate the RPT expressions for the low energy behavior of the model
we need the values of the renormalized parameters in terms of the `bare'
parameters that define the model. We give a brief description how they may be
deduced from the low energy fixed point in an NRG calculation. The results of
these calculations enable us to test the predictions given in Eqns
 (\ref{pred1}) and (\ref{pred2}). An evaluation of the RPT expressions
 gives 
a comprehensive picture of the low energy behavior of the model
in the Fermi liquid regimes.\par

\begin{figure}[!htbp]
   \begin{center}
     \includegraphics[width=0.4\textwidth]{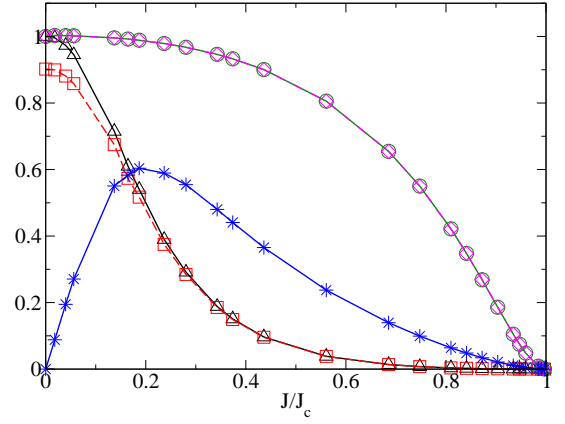}
     \caption{(Color online) A plot of $\tilde\Delta_1(U,J)/\tilde\Delta_1(U,0)$
(dotted line and triangles),
$\tilde\Delta_2(U,J)/\tilde\Delta_2(U,0)$ (full line and circles), $\tilde U_1/\pi\tilde\Delta_1(U,0)$ (dashed
line and squares),
 $\tilde U_2/\pi\tilde\Delta_2(U,0)$ (dot-dashed line and diamonds),
and  $\tilde J/\pi(\tilde\Delta_1(U,0)\tilde\Delta_2(U,0))^{1/2}$ (full line
  and stars)
as a function of $J/J_c$, for  $ U/\pi\Delta_2=6$, $
\Delta_1/\Delta_2=4$, $\Delta_2=0.01$.
}
     \label{asym_rp_scaledU6}
   \end{center}
 \end{figure}

\bigskip
 \begin{figure}[!htbp]
   \begin{center}
     \includegraphics[width=0.4\textwidth]{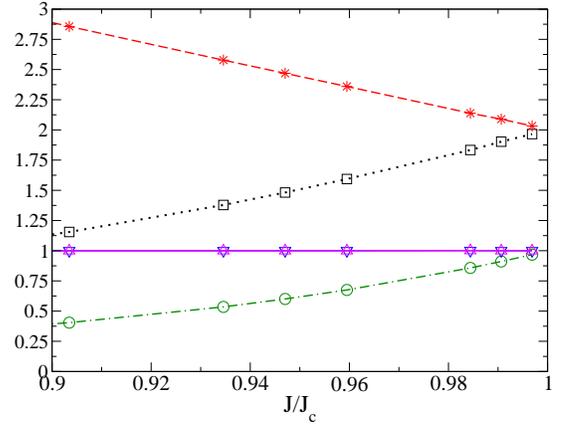}
     \caption{(Color online) A plot of  $\tilde J/\pi\tilde\Delta_1$ (dots and
       squares), 
$\tilde J/\pi\tilde\Delta_2$ (dashes, stars),  $\tilde
U_1/\pi\tilde\Delta_1$(full line, up triangle), 
$\tilde U_2/\pi\tilde\Delta_2$ (full line, down triangle), and   
       $\tilde\Delta_2/\tilde\Delta_1$ (dot-dash, circles) as a function of
       $J/J_c$ for $ U/\pi\Delta_2=6$,   $ \Delta_1/\Delta_2=4$,
         $\pi\Delta_2=0.01$.}
     \label{asym_j_u6}
   \end{center}
 \end{figure}
\begin{figure}[!htbp]
   \begin{center}
     \includegraphics[width=0.4\textwidth]{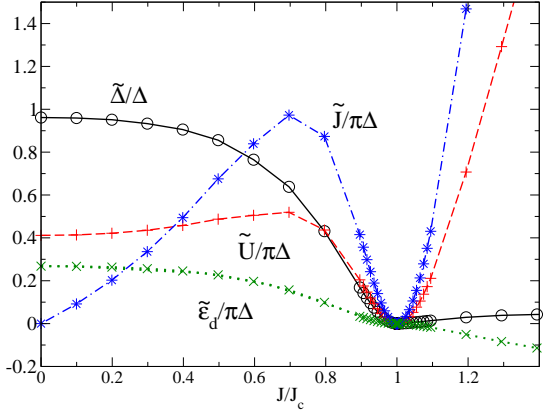}
     \caption{(Color online) A plot of $\tilde\Delta/\Delta$, $\tilde \epsilon_d/\pi\Delta$, 
 $\tilde U/\pi\Delta$ 
and  $\tilde J/\pi\Delta$ 
as a function of $J/J_c$, for  $ U/\pi\Delta=0.5$, $\epsilon_d/\pi\Delta=0.159$.
}
   \label{renpar_ph_asym}
   \end{center}
 \end{figure}
\bigskip
 \begin{figure}[!htbp]
   \begin{center}
     \includegraphics[width=0.4\textwidth]{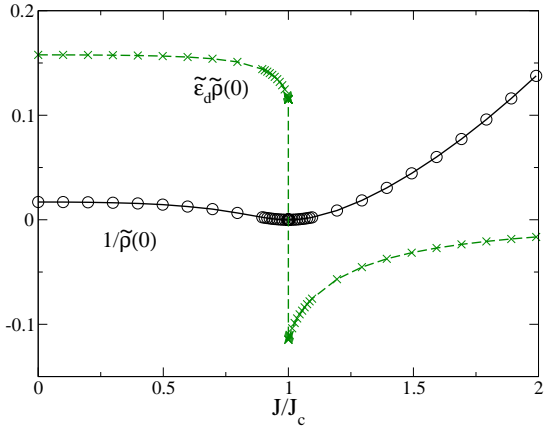}
     \caption{(Color online) A plot of  $1/\tilde\rho(0)$ ($=4T^*$) and
 $\tilde \epsilon_d\tilde\rho(0)$
as a function of $J/J_c$, for  $ U/\pi\Delta=0.5$,
$\epsilon_d/\pi\Delta=0.159$. The discontinuity in   $\tilde
\epsilon_d\tilde\rho(0)$ at the quantum critical point corresponds to a
jump in phase shift of $\pi/2$ per spin state per dot channel. 
}
     \label{nonph_ed}
   \end{center}
 \end{figure}

 \begin{figure}[!htbp]
   \begin{center}
     \includegraphics[width=0.4\textwidth]{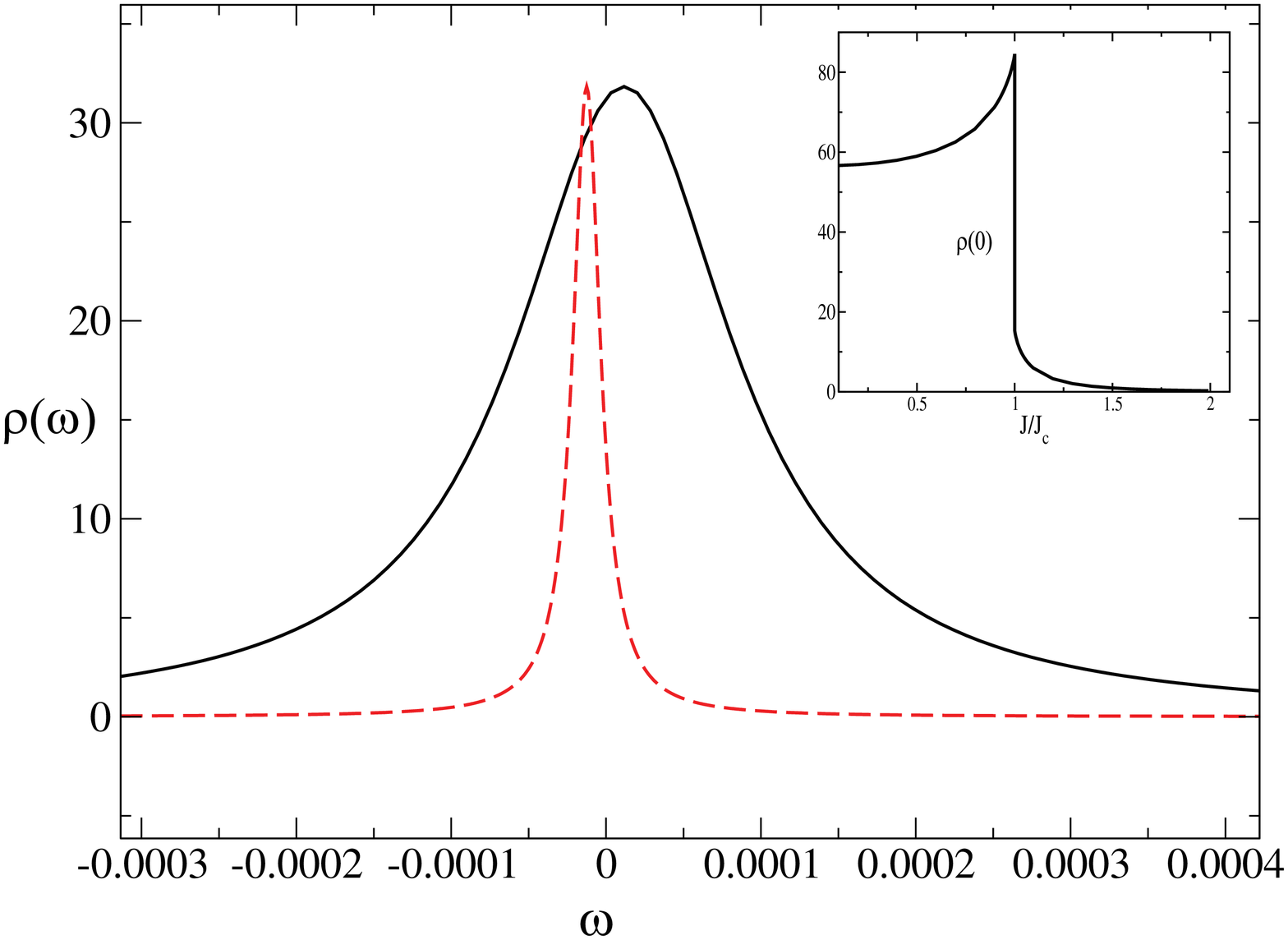}
     \caption{(Color online) A plot of low energy local dot spectral density $\rho(\omega)=z\tilde\rho(\omega)$ 
as a function of $\omega$ for  the two cases $J/J_c=0.98$ (full curve)
and $J/J_c=1.02$ (dashed curve) for the same parameter set as in
Fig. \ref{nonph_ed}. The inset shows $\rho(0)$ ($=z\tilde\rho(0)$) as a
function of $J/J_c$ showing the discontinuous loss of spectral density at the Fermi level
as the value of $J$ increases through the critical point $J=J_c$.
}
     \label{qpdos3}
   \end{center}
 \end{figure}

\bigskip
 \begin{figure}[!htbp]
   \begin{center}
     \includegraphics[width=0.4\textwidth]{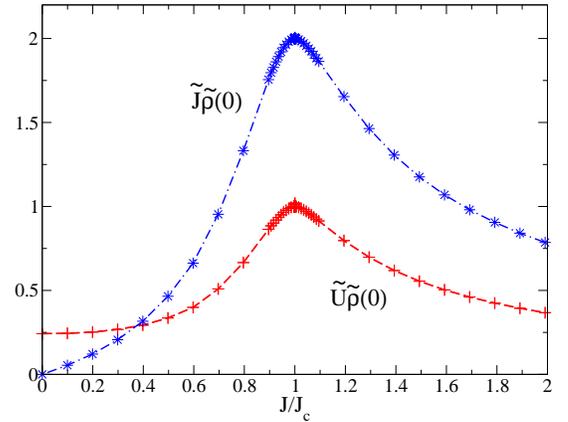}
     \caption{(Color online) A plot of 
 $\tilde U\tilde\rho(0)$ and $\tilde J\tilde\rho(0)$
as a function of $J/J_c$, for  $ U/\pi\Delta=0.5$, $\epsilon_d/\pi\Delta=0.159$.}
     \label{rp_ph_asym}
   \end{center}
 \end{figure}

\section{NRG Calculation of Renormalized Parameters}

The numerical renormalization group results can be used to calculate the
renormalized parameters, by identifying the quasiparticle Hamiltonian in
Eqn. (\ref{qpmodel2c}) as the NRG low energy fixed point with the leading
order
correction terms. 
In the NRG calculation  the non-interacting Green's function for the impurity
site $\alpha$
takes the form,
\begin{equation}
G^{(0)}_{d,\alpha,\sigma}(\omega)={1\over \omega-\epsilon_d-V^2G_{0,\alpha,\sigma}^{(0)}(\omega)}
\end{equation} 
where
\begin{equation}
G^{(0)}_{n,\alpha,\sigma}(\omega)={1\over
  \omega-\epsilon_i-V_n^2G_{n+1,\alpha,\sigma}^{(0)}(\omega)},
\label{cegf}
\end{equation}
with $n=0,1,2,....N$ for a tight binding $N$-site conduction chain, where
$V_n$ are the intersite hopping matrix elements and $\epsilon_n$ the energies,
 with $V_n=\Lambda^{-n/2}\xi_n$, where $\Lambda$ (>1) is the discretization
 parameter, and $\xi_n$ is given by
\begin{equation} \xi_n={D\over 2}{(1+\Lambda^{-1})(1-\Lambda^{-n-1})\over
(1-\Lambda^{-2n-1})^{1/2}(1-\Lambda^{-2n-3})^{1/2}}.
\end{equation}
We use a discretization parameter $\Lambda=6$, $D=1$, particle-hole symmetry in the
conduction band $\epsilon_n=0$,  and retain
4000 states at each NRG iteration.
\par
The renormalized parameters,
 $\tilde\epsilon_d$ and  
$\tilde V$, are calculated by requiring  the lowest energy single  and hole
 particle excitations, $E_{p,\alpha}(N)$ and $E_{h,\alpha}(N)$,  for the {\em
 interacting} system with a NRG chain of length $N$ to correspond to the poles
of the quasiparticle Green's function in the limit of large  $N$.
This gives to the condition, 
\begin{equation}
 E_{p/h,\alpha}(N)-\tilde\epsilon_d-\tilde V^2G_{0,\alpha,\sigma}^{(0)}(E_{p/h,\alpha}(N))=0,\label{rp_eqn1}
\end{equation}  
where $G_{0,\alpha,\sigma}^{(0)}(\omega)$ is given by Eq. (\ref{cegf}).
This equation defines  quantities, $\tilde\epsilon_{d,\alpha}(N)$ and
$\tilde\Delta_{\alpha}(N)$, which in general depend upon $N$. Plotting these
quantities as a function of $N$ a plateau develops when they become
independent
of $N$ for large $N$, which  then determines the parameters
for the free quasiparticles. After diagonalizing the free quasiparticle
Hamiltonian, the renormalized interaction parameters, $\tilde U$, $\tilde J$
and $\tilde U_{12}$, can be calculated
from the leading order correction terms to the difference between
the lowest two particle excitation, $E^{\sigma}_{pp,\alpha}(N)$ and the two lowest
single particle excitations for large $N$, eg. $E^0_{pp,\alpha}(N)-E^{\uparrow}_{p,\alpha}
(N)-E^{\downarrow}_{p,\alpha}(N)$ in channel $\alpha$ determines $\tilde
  U_\alpha$. Once the
renormalized parameters have been determined the $T=0$ susceptibilities
 and specific heat
coefficient $\gamma$ etc  can then be determined by substituting  them into the relevant
RPT equations.
 For further details we refer to earlier papers \cite{HOM04,NCH10s,NCH10a}. 
This analysis differs from the one used by Krishnamurthy, Wilkins and Wilson \cite{KWW80a,KWW80b}
for the Anderson model where a distinction is made between a strong
coupling and weak coupling low energy fixed point.
 In our analysis there is just one type
of low energy fixed point for the Anderson model, the Fermi liquid one.  The relation
between the two approaches is discussed more fully in the Appendix. 
\par

 We now look in detail at the results for the renormalized parameters 
 as the interactions between the dots
are increased from zero to their critical values at the onset of the transitions.\par 
 
\bigskip
 \begin{figure}[!htbp]
   \begin{center}
     \includegraphics[width=0.4\textwidth]{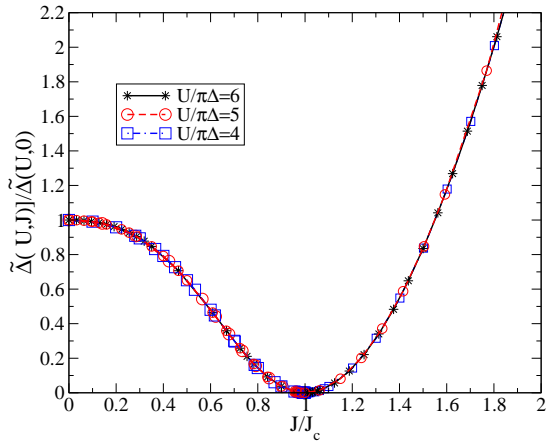}
     \caption{(Color online) A plot of  $\tilde
       \Delta(U,J)/\tilde\Delta(U,0)$ as a function of $J/J_c$ for values of
       $ U/\pi\Delta=4,5,6$. The results fall on a  single curve indicating that
        this ratio becomes universal in the large $U$ regime.}
     \label{delratio1}
   \end{center}
 \end{figure}

\subsection{NRG Results for $U_{12}=0$,  $J<J_c$}

We first of all show results of a calculation for the renormalized
parameters for the symmetric model. Results for
$U/\pi\Delta=5$, which corresponds to the Kondo regime of the model, were given in earlier work \cite{NCH12}, 
so here we show results in a contrasting   regime with
$U=0$   in Fig. \ref{rp_U_0.0}.  Though $U=0$, a
 renormalized interaction $\tilde U$ is induced when the interaction $J$ is
 switched on; though rather small initially, it increases until $J\approx 0.7
 J_c$ and then falls  to zero as  $J\to J_c$. It can be seen  that
all the renormalized parameters, $\pi\tilde\Delta$, $\tilde U$ and $\tilde J$
 tend to zero on the approach to the quantum critical point $J=J_c$. The
 ratios, $\tilde U/\pi\tilde\Delta$ and
$\tilde J/\pi\tilde\Delta$, are shown in Fig. \ref{rp_ratios_u_0}. The value
of $\tilde U/\pi\tilde\Delta$  increases monotonically with increase of $J$
and  tends to the strong coupling value $\tilde U/\pi\tilde\Delta=1$
  on the approach to the critical
point. The ratio $\tilde J/\pi\tilde\Delta$ also increases monotonically such
that  
$\tilde J/\pi\tilde\Delta\to 2$ as $J\to J_c$. These results are in line with the RPT predictions in
Eqn. (\ref{pred1}). \par

We next look at a case where the two dots are
hybridized  differently to their respective baths so we have channel asymmetry.
We take  $V_1=2V_2$, which is such that $\Delta_1=4\Delta_2$, and take
$\Delta_2=0.01$ and $U/\pi\Delta_2=6$. For $J=0$ we are dealing with two
isolated Anderson models where the ratio of the renormalized energy scales of the two
dots, $\tilde\Delta_1(J=0)/
\tilde\Delta_2(J=0)=619.16$, are very different.
In Fig. \ref{asym_rp_scaledU6} we give plots of  $\tilde\Delta_1(U,J)/\Delta_1(U,0)$,
$\tilde\Delta_2(U,J)/\Delta_2(U,0)$, $\tilde U_1/\pi\Delta_1(U,0)$,
 $\tilde U_2/\pi\Delta_2(U,0)$,
and  $\tilde J/\pi(\tilde\Delta_1(U,0)\tilde\Delta_2(U,0))^{1/2}$
as a function of $J/J_c$.
We see that $\tilde U_2=\pi\tilde \Delta_2(U,0)$ at $J=0$ showing that dot 2 is
initially in the Kondo regime, but the ratio $\tilde U_1/\pi\tilde
\Delta_1(U,0)<1$ so that initially dot 1 is not  quite in the Kondo regime
due to the larger value of the hybridization $\Delta_1(U,0)$.
 As $J$ is increased there is a transition at $J_c/\pi\Delta_2=0.1605$
where all the renormalized
parameters tend to zero.
 The results imply that both  $z_1\to 0$ and $z_2\to 0$, 
corresponding to a loss of the local quasiparticle weight at the Fermi level in both dots  at the critical
point.
 \par

To test the predictions based on the RPT given in the Eqn. (\ref{pred1}) for this case,  we plot the  
 ratios of the renormalized parameters,  $\tilde J/\pi\tilde\Delta_1$, 
$\tilde J/\pi\tilde\Delta_2$,  $\tilde
U_1/\pi\tilde\Delta_1$
$\tilde U_2/\pi\tilde\Delta_2$ and   
       $\tilde\Delta_2/\tilde\Delta_1$  as a function of $J/J_c$
      over the range  $0.9<J/J_c<1$. Over this range we see that $\tilde
U_1/\pi\tilde\Delta_1=\tilde
U_2/\pi\tilde\Delta_2=1$, 
corresponding to strong coupling on both dots. Remarkably the ratio  $\tilde\Delta_2/\tilde\Delta_1$, which was initially very
 small, $1/619.16\approx 0.001615$, 
approaches the strong coupling value 1 at the critical point. 
The two ratios, $\tilde J/\pi\tilde\Delta_1$ and  
$\tilde J/\pi\tilde\Delta_2$, converge  to a common value 2 as $J\to J_c$. 
These results confirm the
predictions given in Eqn. (\ref{pred1}) for the case $U_{12}=0$.
Furthermore, they show rather dramatically the emergence of channel
symmetry as $J$ approaches the critical value $J_c$. 
 \par 

We  found precisely similar results  when we fixed the value of $J$
at $J/\pi\Delta_2=0.06$ and
$U/\pi\Delta_2=6$, and increased the hybridization ratio $r=\Delta_1/\Delta_2$.
A transition point was reached at $r=r_c=2.6732$, where all the
renormalized parameters went to zero, and in the predicted ratios.
 \par
 
The transition in this model is remarkably robust and not only persists
in cases where  there is asymmetry between the two dots, but also 
away from particle-hole symmetry \cite{ZCSV06}.  Before looking at an
example of this type we consider how to calculate the renormalized  parameters for the regime
$J>J_c$.\par

\subsection{Extension of NRG results to $J>J_c$}
\subsubsection{Particle-hole asymmetric case}
For the regime $J>J_c$ there is a difference in the calculation of the
renormalized parameters  for the cases with and without particle-hole symmetry.
In both cases the renormalized hybridization, which couples to dots to their
respective conduction baths, $\tilde V_\alpha\to 0$ as $J\to J_c$
for $J<J_c$.  In the particle-hole asymmetric case the renormalized
hybridization re-emerges in the regime $J>J_c$, which means we can apply
precisely the same analysis to the calculation of the renormalized parameters,
$\tilde\Delta$ and $\tilde\epsilon_d$, 
using Eqn. (\ref{rp_eqn1}),
as described earlier. To see what happens as we pass through the transition we
look at the results for a particular example. \par

In Fig. \ref{renpar_ph_asym} we show the results deduced for the renormalized
 parameters over the range $0<J/J_c<1.2$ ($J_c/\pi\Delta=1.0884$) for the  case 
 with  $U/\pi\Delta=0.5$ with
 $\epsilon_d/\pi\Delta=0.159$, where the level on the dot is initially above
 the Fermi level.  It can be seen that all the renormalized parameters, $\tilde
\Delta$,
$\tilde \epsilon_d$, $\tilde U$ and $\tilde J$ go to zero as $J\to J_c$
both for $J<J_c$ and $J>J_c$. 
 To learn in
 more detail what happens to the quantity $\tilde\epsilon_d$  through the
 quantum critical point we plot in Fig. \ref{nonph_ed} the product
 $\tilde\epsilon_d\tilde\rho(0)$ and $1/\tilde\rho(0)$ as a function of
 $J/J_c$.  There is a discontinuity in $\tilde\epsilon_d\tilde\rho(0)$ at $J=J_c$ which reflects the fact that
 there is a discontinuous change in the phase shift $\delta_\alpha$ of
 $\pi/2$. From the Friedel sum rule (\ref{qpocc}) this implies  that  
the total local occupation number in the ground state for $J>J_c$
has two more electrons than that for $J<J_c$. The local density of states $\rho(\omega)$
at each dot in the low frequency regime corresponds to $z\tilde\rho(\omega)$,
and this is plotted in Fig. \ref{qpdos3} for cases on either side of the transition,
$J/J_c=0.98$ and $J/J_c=1.02$. The peak in $\tilde\rho(\omega)$ has shifted from
 just above the Fermi level for $J/J_c=0.98$ to below for $J/J_c=1.02$.
In the inset of Fig. \ref{qpdos3}  $\rho(0)$ ($=z\tilde\rho(0)$) is plotted as a
 function of $J/J_c$. It can be seen that as the value of $J$ increases
 through
the transition  there is a sudden loss of spectral weight at the Fermi
 level but there is finite spectral weight on both sides of the
 transition.\par

 To check the RPT  predictions given in Eqn. (\ref{pred1}) 
 we have plotted  $\tilde U\tilde\rho(0)$ and  $\tilde J\tilde\rho(0)$  
in  Fig. \ref{rp_ph_asym}. It can be seen that  $\tilde U\tilde\rho(0)\to 1$
and  $\tilde J\tilde\rho(0)\to 2$ as $J\to J_c$ on both sides of the quantum
 critical point in agreement with the RPT
prediction. \par

\subsubsection{Particle-hole symmetric case}

The case of particle-hole symmetry differs from the case we have just
considered as the renormalized hybridization parameter $\tilde V_\alpha$,
calculated from the NRG, does not just go to zero as $J\to J_c$ for $J<J_c$ but remains equal to zero 
for all $J>J_c$.  This implies that we lose {\em all} the spectral weight at the 
transition for the particle-hole symmetric case, and not just {\em some} of the
spectral weight as was the case with particle-hole asymmetry.
At the fixed point  for  $J>J_c$  the two dots are decoupled from the conduction electrons
on the lowest energy scale.  We know, however, for $J>J_c$ the low energy
behavior still corresponds to a Fermi liquid. We conjecture that this Fermi
liquid can be described by a similar set of renormalized parameters and that
the equations for the susceptibility and low energy behavior are still valid.
However, this needs to be clarified.\par

The fact that $z\to 0$  implies that the derivative of the 
the self-energy in the impurity Green's function,
$\Sigma'(\omega)$, diverges at $\omega=0$  such that this self-energy
develops a singularity as $J\to J_c$. The fact that $z=0$ for  $J>J_c$
means that that this singularity persists in the regime $J>J_c$
and the analysis outlined in the previous section breaks down. 
To consider this situation in more detail,  we sum over $k$ and $k'$ in Eqn. (\ref{eom}) to obtain an equation
for the Green's function on the first site of the conduction electron chain in
  the NRG calculation which we denote by $G_{0,\alpha,\sigma}(\omega)$,
\begin{equation}G_{0,\alpha,\sigma}(\omega)=G^{(0)}_{0,\alpha,\sigma}(\omega)+
G^{(0)}_{0,\alpha,\sigma}(\omega)V_\alpha^2G_{d,\alpha,\sigma}(\omega) G^{(0)}_{0,\alpha,\sigma}(\omega).\end{equation}
If we take the conduction band  in the wide band limit, $V_\alpha^2 G^{(0)}_{0,\alpha,\sigma}(\omega)=-i\pi\Delta_\alpha$, this equation becomes
\begin{equation}G_{0,\alpha,\sigma}(\omega)=G^{(0)}_{0,\alpha,\sigma}(\omega)\left(1-i\Delta G_{d,\alpha,\sigma}(\omega)\right).
\label{2site}\end{equation}
If we introduce a self-energy $\Sigma^0_{\alpha,\sigma}(\omega)$
for the Green's function for the first site on the $\alpha$ conduction
electron chain, which will also include the effect of switching on the
hybridization term, Eqn. (\ref{2site}) becomes
\begin{equation}
{1+i\rho_c\Sigma^0_{\alpha,\sigma}(\omega)}={1\over 1-i\Delta G_{d,
\alpha,\sigma}(\omega)}.\label{conself}
\end{equation}
For $J<J_c$ the self-energies on the dot are non-singular and from the Friedel
sum rule  $\epsilon_d+\Sigma_\alpha(0)=0$ corresponding to a phase shift $\delta_\alpha=\pi/2$. 
 We then have $G_{d,\alpha,\sigma}(0)=1/i\Delta_\alpha$  for the case of
 particle-hole symmetry and hence from
 Eqns. (\ref{2site}) and   (\ref{conself}) the
Green's function on the first site of the conduction
 chain is zero, $G_{0,\alpha,\sigma}(0)=0$,  and the self-energy $\Sigma^0_{\alpha,\sigma}(\omega)$ 
 diverges at $\omega=0$. \par
For  $J>J_c$, however,  the dot self-energy develops a divergence at
$\omega=0$, such that the dot Green's function vanishes for $\omega=0$,
and we get a phase shift
$\delta_\alpha=0$. The self-energy for the first conduction site
is now non-singular and from Eqn. (\ref{conself}),  $
\Sigma^0_{\alpha,\sigma}(0)=0$. We can then develop a renormalized
perturbation approach and analysis of the low energy NRG fixed point with
 the first conduction site playing  the role of an effective dot.
 We should, therefore,
be able to describe the Fermi liquid behavior for $J>J_c$ in terms
of renormalized parameters associated with the first site on the
impurity conduction electron chain.\par
For $J<J_c$ we determine the renormalized parameters $\tilde\epsilon_d$ and
$\tilde V$ by requiring the lowest energy particle and hole excitations of the
interacting systems to correspond to poles of the impurity quasiparticle
Green's function. For $J>J_c$ we can adopt the same procedure but replacing
the Green's function of the impurity site by that for the first conduction
electron site, so that parameters for $\tilde\epsilon_0$ and $\tilde V_0$
are required to correspond to solutions of the equation,
\begin{equation}
 E_{p/h,\alpha}(N)-\tilde\epsilon_0-\tilde V_0^2G_{1,\alpha,\sigma}^{(0)}(E_{p/h,\alpha}(N))=0,
\end{equation}  
where  $G_{1,\alpha,\sigma}^{(0)}(\omega)$ is given by Eq. (\ref{cegf}).
The local quasiparticle interaction terms, $\tilde U$, $\tilde J$ and $\tilde U_{12}$, can then be calculated as in
the case $J<J_c$.\par
As the effective impurities    now correspond to the first site on
the 
conduction electron chain for each channel, there are a some modifications
to the analysis used for $J<J_c$. The density of states of the conduction electron chain which now
starts at the second conduction site is modified, which affects the
calculation of the effective hybridization width
$\Delta_0=\pi|V_0|^2\rho_c$, where $\rho_c$ is the density of conduction
states at the Fermi level. The factor $A_\Lambda$, which relates the
hybridization width of the discretized model to the continuum model,  also
changes.  These two effects can be combined into a single additional factor
$B_\Lambda$ so that the hybridization width $\Delta_0$
for the continuum model is calculated from  $\Delta_0=\pi|V_0|^2/2A_\Lambda
B_\Lambda D$. The value of $B_\Lambda$ was estimated from the
result for the isolated Anderson model  ($J=0$) for $U/\pi\Delta\gg 1$,
$\tilde U/\pi\tilde\Delta\to 1$, which for a discretization parameter
 $\Lambda=6$, gave the  value $B_6=2.2373$. \par
In the limit $J\to \infty$ the two impurities are entirely decoupled from the
conduction electrons. This is reflected in the NRG results for the
renormalized parameter  $\tilde V_0$, which in this limit  give  $\tilde V_0\to
\Lambda^{-1/2}\xi_1$, which is the value for the free NRG
conduction chain. The calculation of the impurity susceptibility using the 
RPT expression with $\tilde\Delta_0$ includes this extra conduction electron contribution. However,
this correction  is negligible except in the regime when $J/J_c$ is very large
and 
it is not necessary to take it into account. \par

\begin{figure}[!htbp]
   \begin{center}
     \includegraphics[width=0.4\textwidth]{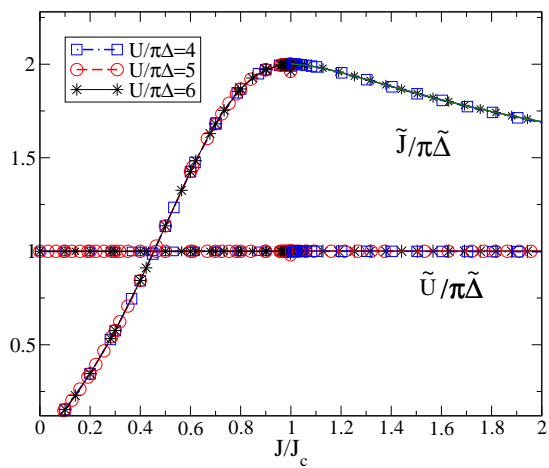}
     \caption{(Color online)A plot of  $\tilde U/\pi\tilde \Delta$  and  $\tilde J/\pi\tilde \Delta$, as a
       function of $J/J_c$ for  $ U/\pi\Delta=4,5,6$.
} 
     \label{ratios456}
   \end{center}
 \end{figure}

\subsection{NRG results for $J\ne 0$, $U_{12}=0$}

We now consider the channel and particle-hole  symmetric model taking $V_1=V_2$ so we can drop
 the channel index $\alpha$. We extend the results for the renormalized
 parameters to include the regime $J>J_c$. We restrict attention first of all to the
strong correlation regime 
$U/\pi\Delta\gg 3$ for different values of $U$.  
In Fig. \ref{delratio1} we plot the results for the ratio
$\tilde\Delta(U,J)/\tilde\Delta(U,0)$ as a function of $J/J_c$ for
$U/\pi\Delta=4,5,6$ over the range $0<J/J_c<2$. It can be seen that, though the values of $J_c$ vary significantly, all three results fall on a single
curve indicating  universal behavior in this regime. We have dropped the index 
 on $\Delta_0$
 and  $\tilde\Delta_0$ for the regime $J>J_c$, to define a continuous
function
  $\tilde\Delta$ through the transition point.  
With the definition of the Kondo temperature $T_{\rm K}$,
 $\pi\tilde\Delta(U,0)=4T_{\rm K}$  and the energy scale 
  $T^*(U,J)$, $\pi\tilde\Delta(U,J)=4T^*$ we now have a single energy
 scale over the range $0 < J/J_c < 2$
such that  in the strong correlation regime,
\begin{equation}
                  T^*=T_{\rm K}F(J/J_c),
\label{univ1}
\end{equation}
where the function $F(J/J_c)$  is independent of $U$ for $U/\pi\Delta\gg
1$, such that $F(0)=1$ and $F(1)=0$. We also have $J_c=1.378T_{\rm K}$ in this
regime, so Eqn. (\ref{univ1}) can be re-expressed in the form, $T^*=T_{\rm
  K}F(0.726J/T_{\rm K})$.
\par

In Fig. \ref{ratios456} we plot the ratios   $\tilde U/\pi\tilde\Delta$
and 
  $\tilde J/\pi\tilde\Delta$, for
 $U/\pi\Delta=4,5,6$ over the same range.
All the values for  $\tilde U/\pi\tilde\Delta$ fall on the same line $\tilde
U/\pi\tilde\Delta=1$ corresponding to the strong coupling limit.
 The values of $\tilde J/\pi\tilde\Delta$ also all lie on a single  curve.
Hence, using
 Eqn. (\ref{univ1}), in this regime we have 
\begin{equation}
{\tilde J}=4T^*f({J/ J_c}),
\label{univ2}
\end{equation}
where $f(J/J_c)$ is a universal function such that $f(0)=0$ and $f(1)=2$.
As from Eqn. (\ref{univ1}) $T^*$ is a universal function of $J$ and $T_{\rm
  K}$
then Eqn. (\ref{univ2}) can be re-expressed as ${\tilde J}=T_{\rm K}\bar f({0.726J/
  T_{\rm K}})$, where $\bar f(J/J_c)$ is a universal function such that
$\bar f(0)=0$ and 
$\bar f(1)=0$.
\par

  \begin{figure}[!htbp]
    \begin{center}
      \includegraphics[width=0.4\textwidth]{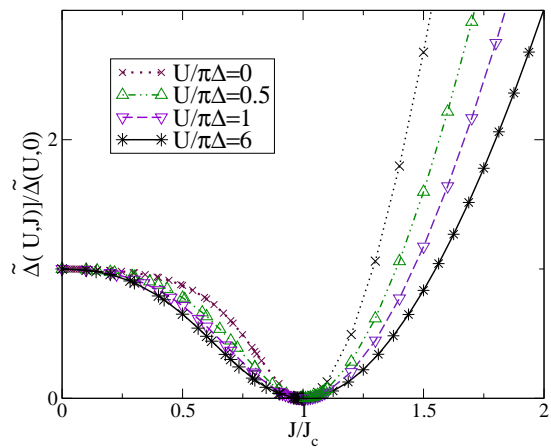}
      \caption{(Color online) A plot of  $\tilde
        \Delta(U,J)/\tilde\Delta(U,0)$ as a function of $J/J_c$ for values of
        $ U/\pi\tilde\Delta=0,0.5,1,6$.}
      \label{delratio2}
    \end{center}
  \end{figure}

\begin{figure}[!htbp]
   \begin{center}
     \includegraphics[width=0.4\textwidth]{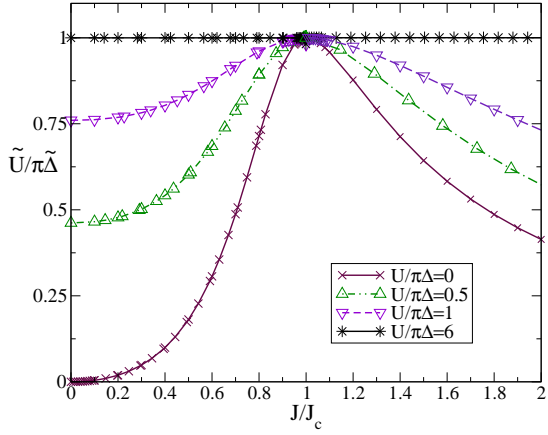}
     \caption{(Color online) A plot of  $\tilde U/\pi\tilde \Delta$, as a
       function of $J/J_c$ for  $ U/\pi\Delta=0,0.5,1,6$.
} 
     \label{tilu}
   \end{center}
 \end{figure}

The universality found in the Kondo regime $U/\pi\Delta\gg 1 $ no longer holds
for smaller values of $U$ as can be seen in
Fig. \ref{delratio2} where we give results for the ratio of $\tilde\Delta(U,J)/\tilde
\Delta(U,0)$ over the range $0<J/J_c<2$ for  values of $U/\pi\Delta=0,0.5,1,2,6$ as a function of
$J/J_c$. It can be seen that they all qualitatively behave in a similar way,
but for smaller values of $U/\pi\Delta$ the results do not lie on the same
curve.\par
\begin{figure}[!htbp]
   \begin{center}
     \includegraphics[width=0.4\textwidth]{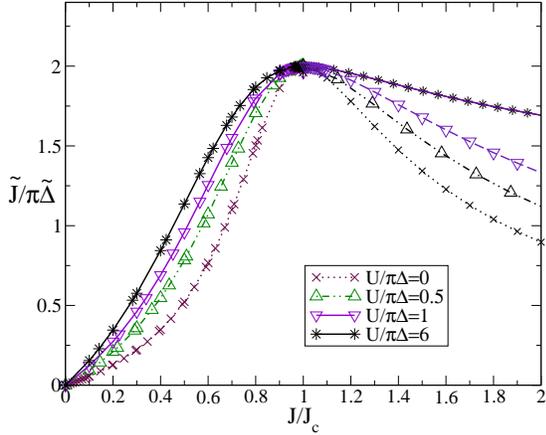}
     \caption{(Color online) A plot of  $\tilde J/\pi\tilde \Delta$, as a
       function of $J/J_c$ for  $ U/\pi\Delta=0,0.5,1, 6$.
} 
     \label{tilj}
   \end{center}
 \end{figure}

  In Figs. \ref{tilu} and  \ref{tilj}
 we give the corresponding  plot for  the ratios $\tilde U/\pi\tilde
\Delta$ and  $\tilde J/\pi\tilde\Delta$. There is no universality  
except in the approach to the critical point where 
 $\tilde U/\pi\tilde \Delta\to 1$ and  $\tilde J/\pi\tilde \Delta\to 2$
in all cases in line with the predictions in Eqn. (\ref{pred1}).\par

\begin{figure}[!htbp]
   \begin{center}
     \includegraphics[width=0.4\textwidth]{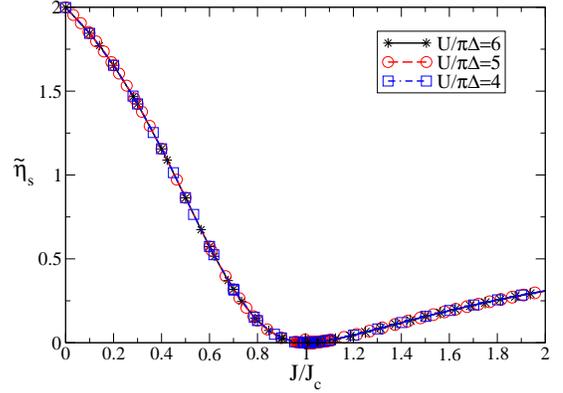}
     \caption{(Color online) A plot of the Wilson ratio  $\tilde\eta_s$ as a
       function of $J/J_c$ for  $ U/\pi\Delta=4,5,6$ for  $\pi\Delta=0.01$.
} 
     \label{etas456}
   \end{center}
 \end{figure}

\begin{figure}[!htbp]
   \begin{center}
     \includegraphics[width=0.4\textwidth]{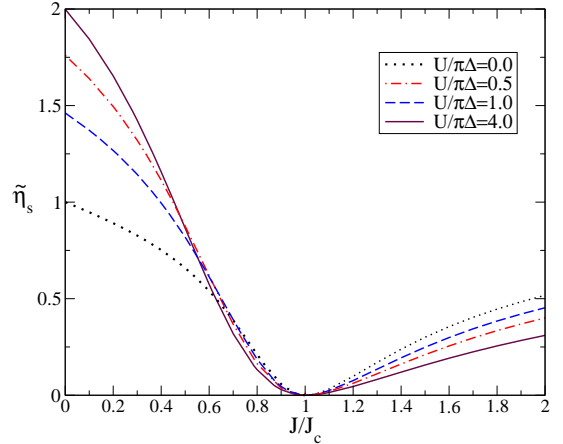}
     \caption{(Color online) A plot of  $\tilde\eta_s$, as a
       function of $J/J_c$ for  $ U/\pi\Delta=0,0.5,1,4$ for  $\pi\Delta=0.01$.
} 
     \label{etas_U_0}
   \end{center}
 \end{figure}

The factors $\tilde\eta_s$ and $\tilde\eta_c$ in the expressions for the spin and
charge susceptibilities reflect the effect of the interaction between the
quasiparticles in enhancing or suppressing the corresponding fluctuations.
For the single impurity Anderson model  $\tilde\eta_s$  is better known as the Wilson ratio and in the
strong coupling regime has the value 2, enhanced over that for non-interacting
quasiparticles where the value would be 1. In Fig. \ref{etas456} we plot
the values of $\tilde\eta_s$ for $U/\pi\Delta=4,5,6$ as a function of 
$J/J_c$. We see that there is a universal curve in this regime which falls
from the value 2 for the isolated impurities to zero at the critical point.
 In Fig. \ref{etas_U_0} the values of $\tilde\eta_s$
are plotted as a function of $J/J_c$ for  $U/\pi\Delta=0,0.5,1,4$. 
There is non-universal behavior in the regime, though the results for
$U/\pi\Delta=4$ correspond to the universal curve for the strong coupling limit.
\bigskip

\begin{figure}[!htbp]
   \begin{center}
     \includegraphics[width=0.4\textwidth]{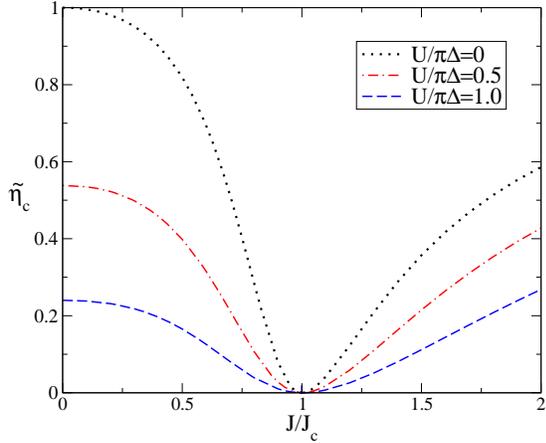}
     \caption{(Color online) A plot of  $\tilde \eta_c$, as a
       function of $J/J_c$ for  $ U/\pi\Delta=0,0.5,1$ for  $\pi\Delta=0.01$.
} 
     \label{etac_U_0}
   \end{center}
 \end{figure}

The values of $\tilde\eta_c$ for $U/\pi\Delta=0,0.5,1$
are shown in Fig. \ref{etac_U_0}. For $U=0$ and $J=0$ the value is 1
corresponding to a non-interacting system and  then, as $J$ is
increased,  progressively reduces to
zero (corresponding to the Kondo limit) as $J\to J_c$.  For the larger values of $U$,
 $\tilde\eta_c$ is already reduced at $J=0$, due to the
on-site interaction $U$, and is then further reduced to zero as the exchange
interaction increases to the critical value. For $U/\pi\Delta>3.5$,
$\tilde\eta_c\approx 0$ as the charge fluctuations are almost completely suppressed  for all values of $J$ in this range.\par

\begin{figure}[!htbp]
   \begin{center}
     \includegraphics[width=0.4\textwidth]{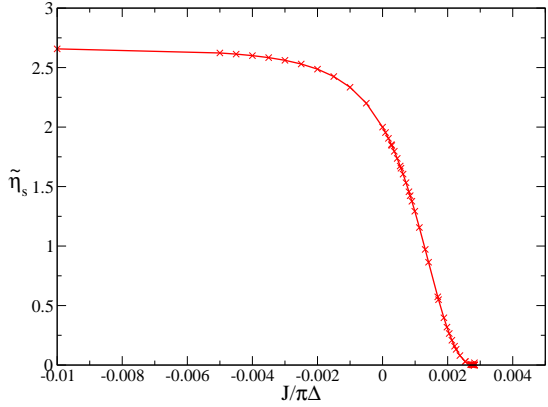}
     \caption{(Color online) A plot of  $\tilde\eta_s$ (Wilson ratio) as a
       function of $J/\pi\Delta$, including the ferromagnetic range,  for  $U/\pi\Delta=5$ for  $\pi\Delta=0.01$.
} 
     \label{ext_etas_U5}
   \end{center}
 \end{figure}

All the expressions given in the previous section apply equally well to the
situation with a ferromagnetic exchange $J<0$. In figure \ref{ext_etas_U5}
we show the values of $\tilde\eta_s$ as a function of $J/\pi\Delta$ extended
to the ferromagnetic range for $U/\pi\Delta=5$, where it extrapolates to the
expected value $8/3$.

\subsection{Results for $U_{12}< U_{12}^c$, $J=0$}

The transition to a locally charge ordered state in the model 
with a finite interaction $U_{12}$ ($J=0$) has been studied in detail
by Galpin et al \cite{GLK06}. Here we throw further light on this transition by
calculating the renormalized parameters. In particular we test the RPT
conjecture
of the emergence of a single energy scale on the approach to the
QCP and the  predictions given in Eqn. (\ref{pred2}).\par

\begin{figure}[!htbp]
   \begin{center}
     \includegraphics[width=0.4\textwidth]{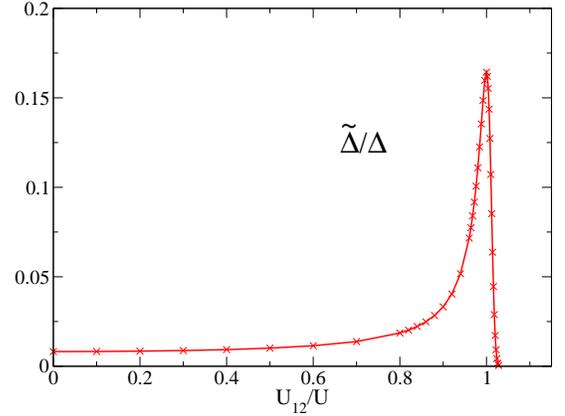}
     \caption{(Color online) A plot of $ \tilde\Delta/\Delta$
as a function of $U_{12}/U$ for  $U/\pi\Delta=5$.}
     \label{tildedelU12}
   \end{center}
 \end{figure}

\begin{figure}[!htbp]
   \begin{center}
     \includegraphics[width=0.4\textwidth]{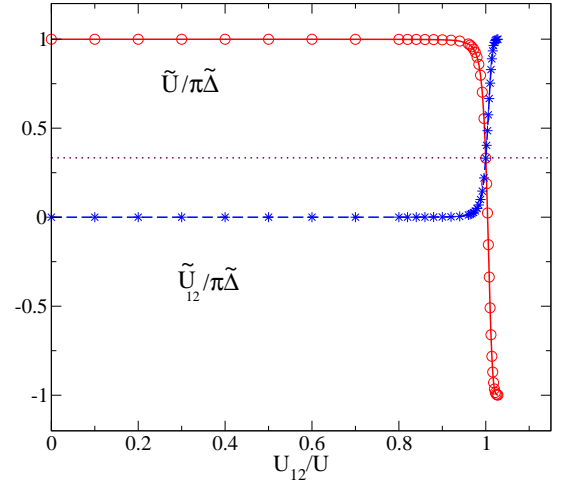}
     \caption{(Color online) A plot of $\tilde U/\pi\tilde\Delta$ (circles)
and $\tilde U_{12} /\pi\tilde\Delta$ (stars)
as a function of $U_{12}/U$ for  $U/\pi\Delta=5$. The dotted line which passes through the SU(4) point
corresponds to  $\tilde U_{12}/\pi\tilde\Delta= \tilde U/\pi\tilde\Delta=1/3$.}
     \label{rp_ratiosU12}
   \end{center}
 \end{figure}

In Fig. \ref{tildedelU12} we give results for the ratio $\tilde\Delta/\Delta$
as a function of $U_{12}/U$ for the case $U/\pi\Delta=5$. Over the range
$U_{12}/U<0.8$ there is a slow steady increase in $\tilde\Delta$ with
increasing
$U_{12}$. There is then a rapid increase due to fluctuations of
charge between the two dots and $\tilde\Delta$ reaches a maximum
at the SU(4) point $U_{12}=U$, and an extremely rapid fall off beyond
this point to a QCP at  $U_{12}/U=1.028$. In Fig. \ref{rp_ratiosU12}
the corresponding  results for the renormalized parameter ratios, $\tilde U/ \pi\tilde\Delta$ 
and $\tilde U_{12} /\pi\tilde\Delta$, are shown as a function of $U_{12}/U$
over the same range. Over the range $U_{12}/U<0.9$ these ratios differ
very little from the values for the Anderson model in the Kondo regime,
 $\tilde U/ \pi\tilde\Delta=1$ and $\tilde U_{12} /\pi\tilde\Delta=0$. 
Beyond this point there is a rapid change and the two curves cross
at the SU(4) point where  $\tilde U/ \pi\tilde\Delta=\tilde U_{12}
/\pi\tilde\Delta ={1/3}$. This result follows from the RPT equations
for the susceptibility from the condition that the  uniform charge
susceptibility $\chi_c$ is so small, due to the large value of $U$,
that it can be equated to zero. Beyond the SU(4) point the ratios
rapidly approach the  values predicted in Eqn. (\ref{pred2})  on the approach to the QCP,
  $\tilde U/ \pi\tilde\Delta=-1$ and $\tilde U_{12} /\pi\tilde\Delta=1$. \par
\bigskip  
\bigskip
 
 \noindent
 \noindent
 \noindent

\begin{figure}[!htbp]
   \begin{center}
     \includegraphics[width=0.4\textwidth]{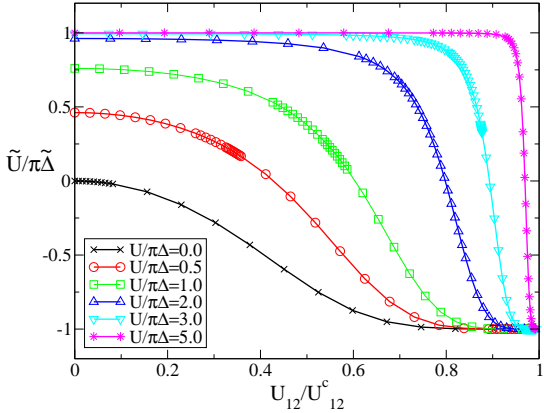}
     \caption{(Color online) Plots of $\tilde U/\pi \tilde\Delta$ 
as a function of $U_{12}/U_{12}^c$ for values of $U/\pi\Delta=0,0.5,1,2,3,5$, $J=0$.}
     \label{ru_vs_u12}
   \end{center}
 \end{figure}

The question arises as to what happens for smaller values of $U$ and in
particular
the case $U=0$. This is addressed in Fig. \ref{ru_vs_u12} which shows a plot
of $\tilde U /\pi\tilde\Delta$ as a function of $U_{12}/U_{12}^c$ for values 
of $U/\pi\Delta=0,0.5,1,2,3,5$. We see that in all cases $\tilde U
/\pi\tilde\Delta\to -1$ as $U_{12}\to U_{12}^c$, though less rapidly the
smaller
the value of $U$. This also holds for the case $U=0$ so that switching on
the interaction term $U_{12}$ induces an onsite effective interaction
$\tilde U$ in the quasiparticle Hamiltonian.\par
\begin{figure}[!htbp]
   \begin{center}
     \includegraphics[width=0.4\textwidth]{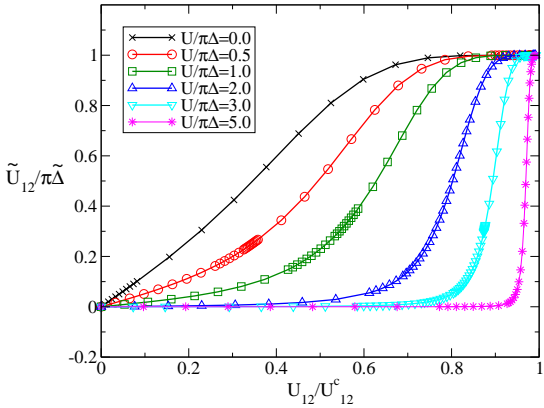}
     \caption{(Color online) Plots of $\tilde U_{12}/\pi \tilde\Delta$ 
as a function of $U_{12}/U_{12}^c$ for values of $U/\pi\Delta=0,0.5,1,2,3,5$, $J=0$.}

     \label{ru12_vs_u12}
   \end{center}
 \end{figure}

The corresponding curves for the ratio  $\tilde U_{12}/\pi \tilde\Delta$ 
as a function of $U_{12}/U_{12}^c$ are shown in Fig. \ref{ru12_vs_u12}.
We see that it this case all the curves approach the predicted value
$\tilde U_{12}/\pi \tilde\Delta\to 1$   as $U_{12}\to U_{12}^c$,
though much less rapidly for the smaller values of $U$.\par

 \noindent

 \noindent
 \noindent

\subsection{ NRG results for $J\ne 0$ and $U_{12}\ne 0$}

\begin{figure}[!htbp]
   \begin{center}
     \includegraphics[width=0.4\textwidth]{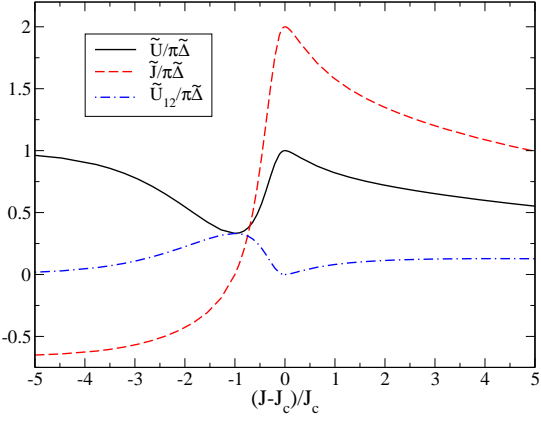}
     \caption{(Color online) A plot of the renormalized parameter ratios, $\tilde J/\pi\tilde\Delta$, $\tilde U/\pi\tilde\Delta$ and $\tilde U_{12}/\pi\tilde\Delta$ as a function of $(J-J_c)/J_c$ where
 $J_c=0.09015\pi\Delta$ for  $U/\pi\Delta=U_{12}/\pi\Delta=5$. As $J$ is increased  there is an SU(4) point at  $J=0$ where  $\tilde U/\pi\tilde\Delta=\tilde U_{12}/\pi\tilde\Delta=1/3$ and  $\tilde J=0$  followed by a transition at $J=J_c$ from the Fermi liquid
phase $J<J_c$ to the second Fermi liquid phase $J>J_c$. }
     \label{u5_v5_vs_j}
   \end{center}
 \end{figure}

To complete the check on the predictions in Eqns. (\ref{pred1}) and
(\ref{pred2}) we look at some examples with finite values of both
$J$ and $U_{12}$. We look first at a case with
$U/\pi\Delta=U_{12}/\pi\Delta=5$ over a range of values of $J$ through
the critical point $J_c=0.09015\pi\Delta$. This corresponds to moving along 
a vertical line in Fig. \ref{phased2} with $U_{12}/D=0.05$.
The results for the renormalized parameter ratios,  $\tilde
J/\pi\tilde\Delta$, $\tilde U/\pi\tilde\Delta$ and $\tilde
U_{12}/\pi\tilde\Delta$ are shown in Fig. \ref{u5_v5_vs_j}.
The results include the regime with ferromagnetic coupling $J<0$,
and it can be seen that as the  ferromagnetic coupling increases
$\tilde U/\pi\Delta$ approaches the value 1 and  $\tilde
J/\pi\tilde\Delta$ the value $-2/3$ as expected in the regime
where the channel and charge fluctuations are suppressed.
The value of $U_{12}/\pi\tilde\Delta$ is small in this limit
but increases as the ferromagnetic coupling is reduced. 
At $J=0$   there is an SU(4) point
where the curves for  $\tilde U/\pi\tilde\Delta$ and $\tilde
U_{12}/\pi\tilde\Delta$ meet at a common value $1/3$, and
$\tilde J=0$. As $J$ is increased from $J=0$ the  curves for  $\tilde
J/\pi\tilde\Delta$ and $\tilde U/\pi\tilde\Delta$ increase to a maximum 
at the transition point $J=J_c$, where they reach the values
2 and 1, respectively. At this point $\tilde U_{12}/\pi\tilde\Delta=0$
in line with the general prediction in Eqn. (\ref{pred1}).
This means that switching on and increasing the value of $J$ 
have the effect of cancelling out the $\tilde U_{12}/\pi\Delta$ ratio
 which is finite for $J=0$, and reducing it to zero as $J\to J_c$.
For $J>J_c$, the values of  $\tilde
J/\pi\tilde\Delta$, $\tilde U/\pi\tilde\Delta$ fall off significantly
with increase of $J$ and $\tilde U_{12}/\pi\tilde\Delta$ has a slow increase.\par
\begin{figure}[!htbp]
   \begin{center}
     \includegraphics[width=0.4\textwidth]{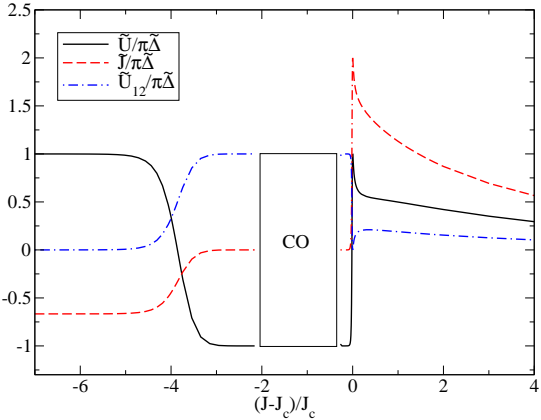}
     \caption{(Color online)  A plot of the renormalized parameter ratios, $\tilde J/\pi\tilde\Delta$, $\tilde U/\pi\tilde\Delta$ and $\tilde U_{12}/\pi\tilde\Delta$ as a function of $(J-J_c)/J_c$ where
 $J_c=0.27228\pi\Delta$ for  $U/\pi\Delta=5$, $U_{12}/\pi\Delta=5.2$. As $J$ is increased there  is first transition from a Fermi liquid phase $J<J_c$ to  a local charged ordered regime indicated by the box (CO) followed by a return to the Fermi liquid
phase $J<J_c$ and then a transition to the second Fermi liquid phase $J>J_c$.}
     \label{u5_v5.2_vs_j}
   \end{center}
 \end{figure}

 In Fig. \ref{u5_v5.2_vs_j} we plot the same ratios of the renormalized
parameters but in this case we take value $U_{12}/\pi\Delta=5.2$.
We take a range of values of $J$ corresponding to moving along a vertical
line in Fig. \ref{phased2}, $U_{12}/D=0.052$. We see in this case the line passes through
three QCPs. For the ferromagnetic regime where $(J-J_c)/J_c<-6$, the spin and
inter-dot fluctuations are suppressed and the ratios take on the expected
values, $\tilde U/\pi\tilde\Delta=1$, $\tilde
J/\pi\tilde\Delta =-2/3$ and $\tilde U_{12}/\pi\tilde\Delta=0$.
Around $(J-J_c)/J_c\sim-4$ there is a rapid change over as the values
move to $\tilde U/\pi\tilde\Delta=-1$, $\tilde
J/\pi\tilde\Delta =0$ and $\tilde U_{12}/\pi\tilde\Delta=1$
on the approach to the local charge order transition at  $(J-J_c)/J_c\approx
-2.1$.  When  $(J-J_c)/J_c\approx
-0.25$ the system emerges from the charge ordered state, with the same set of ratios,
but as $J\to J_c$ there is a rapid change of the ratios of the renormalized
parameters to the ratios, $\tilde U/\pi\tilde\Delta=1$, $\tilde
J/\pi\tilde\Delta =2$ and $\tilde U_{12}/\pi\tilde\Delta=0$ on the approach to
the transition to the local singlet state. There is a rapid decrease in
$\tilde U/\pi\tilde\Delta$, $\tilde
J/\pi\tilde\Delta$ and an increase in the ratio $\tilde
U_{12}/\pi\tilde\Delta$.
For  $(J-J_c)/J_c>0.25$ there is a slow fall off with increase of $J$ for all
three ratios.\par

 \noindent

 \noindent
 \noindent

\subsection{The case $U_{12}=U+3J/2$}

There is a special line in the phase diagram shown in Fig. \ref{phased2},
corresponding to $U_{12}=U+3J/2$ along which there appears to be no transition.
With this  set of parameters, and a ferromagnetic
exchange coupling ($J<0$), the model corresponds to a model introduced
by Yoshimori \cite{Yos76} to describe a single impurity with an $n$-fold
degenerate orbital
 for the case $n=2$. 
 The exchange interaction $J$ in this context corresponds to a Hund's rule
term.  
 The model in the
ferromagnetic regime was used to calculate the orbital susceptibility\cite{NCH10s,NCH10a}
for a two-fold degenerate impurity with a Hund's rule interaction $J$,
where the orbital susceptibility is given by  
\begin{equation}
\chi_{orb}={\mu_{\rm B}^2\eta_{orb}\tilde \rho(0)\over 4}, \quad 
\eta_{orb}=1+(\tilde U+3\tilde J)\tilde\rho(0).\label{chiorb}
\end{equation}
As the value of $|J|$ is increased in the
ferromagnetic range the orbital fluctuations are suppressed and the two local
spins are coupled to give an effective spin 1. When $\tilde\eta_{\rm orb}\sim
0$ from Eq. (\ref{chiorb}) we find the value, 
$|\tilde J|=2\pi\tilde\Delta/3$.  The
 value  $\tilde\eta_s$  approaches  $\tilde\eta_s=8/3$, as expected for
the Wilson ratio for a two channel Kondo model.\par 

We have previously calculated the renormalized
parameters for this model, both for the particle-hole symmetric and
antisymmetric models, for a ferromagnetic coupling   \cite{NCH10s,NCH10a}.
It is interesting to check here what happens in the antiferromagnetic case.
The lack of a transition is likely to be due to the fact that for the isolated
dots at half filling the gain in energy in forming a local singlet $-3J/2$ is
compensated by an equal and opposite term from the $J$-dependence of the
$U_{12}$ term. We look at the results for the renormalized parameters in detail.
\par

 \noindent
\begin{figure}[!htbp]
   \begin{center}
     \includegraphics[width=0.4\textwidth]{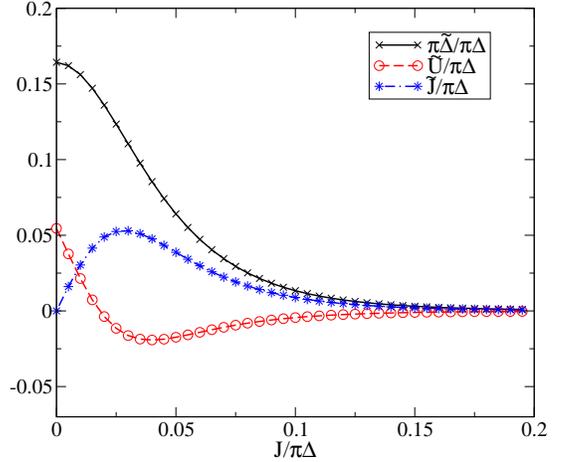}
     \caption{(Color online) A plot of the renormalized
       parameters,$\pi\tilde\Delta/\pi\Delta$, $\tilde U/\pi\Delta$ and
       $\tilde J/\pi\Delta$ as a function of
 $J/\pi\Delta$ for $U/\pi\Delta=5$, $U_{12}=U+3J/2$.}
     \label{rp_hund}
   \end{center}
 \end{figure}
 \noindent

 \noindent
\begin{figure}[!htbp]
   \begin{center}
     \includegraphics[width=0.4\textwidth]{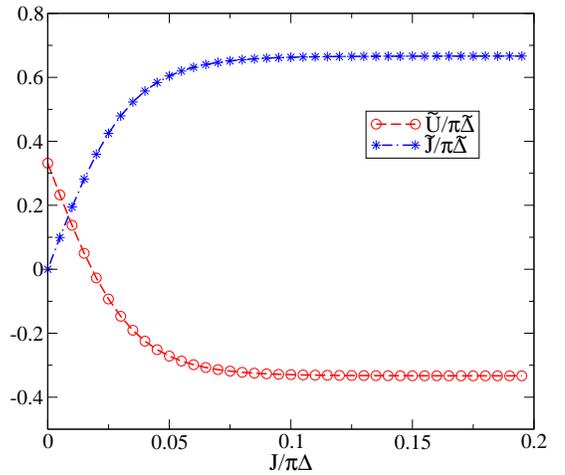}
     \caption{(Color online) A plot of ratios of the renormalized
       parameters, $\tilde U/\pi\tilde \Delta$ and
       $\tilde J/\pi\tilde \Delta$ as a function of
 $J/\pi\Delta$ for $U/\pi\Delta=5$, $U_{12}=U+3J/2$.}
     \label{ratios_hund}
   \end{center}
 \end{figure}
 \noindent

 In Fig. \ref{rp_hund} we plot the renormalized parameters
$\pi\tilde\Delta/\pi\Delta$, $\tilde U/\pi\Delta$ and  $\tilde J/\pi\Delta$
as a function of  $J/\pi\Delta$ over an antiferromagnetic range of $J$
for  $ U/\pi\Delta=5$.
 The value of $\tilde \Delta$, and consequently the
quasiparticle
weight factor $z$, decreases monotonically to very small values for large $J$
without any evidence of a transition. Somewhat surprisingly in this case
the value of  $\tilde U/\pi\Delta$ decreases and becomes negative with
increase
of $J$.\par

 In Fig. \ref{ratios_hund} we plot that ratios  $\tilde U/\pi\tilde \Delta$ and
       $\tilde J/\pi\tilde \Delta$ as a function of
 $J/\pi\Delta$. For $J=0$, $\tilde U/\pi\tilde\Delta=1/3$, 
which is predicted from the condition that the $\tilde\eta_c\sim 0$
due to the large value of $U/\pi\Delta$ which suppresses the charge
susceptibility. For  large $J$,   $\tilde
 U/\pi\tilde \Delta\to -1/3$ and
       $\tilde J/\pi\tilde \Delta\to 2/3$, so there is a single energy scale
       for large values of $J$. These results are in line with  predictions based on
       the fact that for large $U$ and large $J$ both the charge and spin
       susceptibilities are suppressed so $\tilde\eta_c\to 0$ and $\tilde \eta_s\to 0$.
 From Eq. (\ref{etas}) the condition  $\tilde\eta_c\to 0$
  we find $3(\tilde U+\tilde J)=\pi\tilde\Delta$.
Using this result in the expressions for $\tilde\eta_s$ and $\tilde\eta_{orb}$
we find
\begin{equation} \tilde\eta_s={4\over 3}-{2\tilde J\over\pi\tilde\Delta},\quad
  \tilde\eta_{orb}={4\over 3}+{2\tilde J\over\pi\tilde\Delta}.\label{etasorb}\end{equation}
From the condition $\tilde \eta_s\to 0$, using Eq. (\ref{etas}), we find  $\tilde J/\pi\tilde\Delta\to
2/3$.  When this result is substituted into  the  earlier condition,
$3(\tilde U+\tilde J)=\pi\tilde\Delta$, we find
  $\tilde U/\pi\tilde\Delta\to
-1/3$ in this limit, and  $\tilde\eta_{orb}\to 8/3$.\par

\begin{figure}[!htbp]
   \begin{center}
     \includegraphics[width=0.4\textwidth]{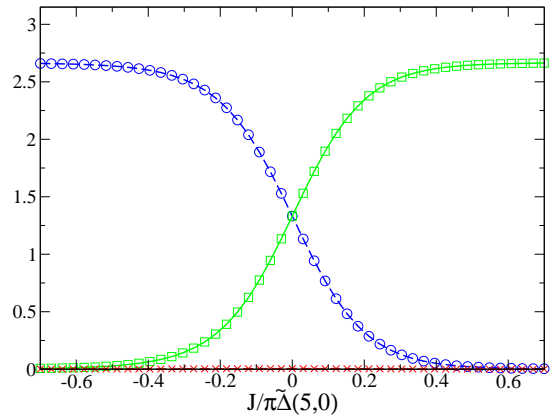}
     \caption{(Color online) A plot of $\tilde \eta_s$ (circles),
 $\tilde \eta_{orb}$ (squares) and $\tilde \eta_c$
(crosses)  as a function of
 $J/\pi\tilde\Delta(5,0)$ for $U_{12}=U+3J/2$, $U/\pi\Delta=5$.}
     \label{eta_hundmodel}
   \end{center}
 \end{figure}

From  earlier results for this model in the ferromagnetic range \cite{NCH10s}, when
 $|J|$ was large such that the orbital susceptibility was suppressed we 
found $\tilde J/\pi\tilde\Delta\to -2/3$ and $\tilde\eta_s\to 8/3$, which
 suggests from Eq. (\ref{etasorb}) the relation $\tilde\eta_s\leftrightarrow
 \tilde\eta_{orb}$ if $\tilde J/\pi\tilde\Delta\leftrightarrow -\tilde J/\pi\tilde\Delta$
under the sign change $J\to -J$. In Fig. \ref{eta_hundmodel} we plot
the values of  $\tilde \eta_s$, $\tilde \eta_{orb}$ and $\tilde \eta_c$
as a function of $J/\tilde\Delta(5,0)$ for $U/\pi\Delta=5$. The results show
 that there is an approximate symmetry  $\tilde\eta_s\leftrightarrow
 \tilde\eta_{orb}$ as  $J\leftrightarrow -J$, but it is not precise for
values in the intermediate range  of $|J|$ where the differences, though small, are greater
than any that might arise from the estimated errors in the calculation of the
 renormalized parameters.
\par
\begin{figure}[!htbp]
   \begin{center}
     \includegraphics[width=0.4\textwidth]{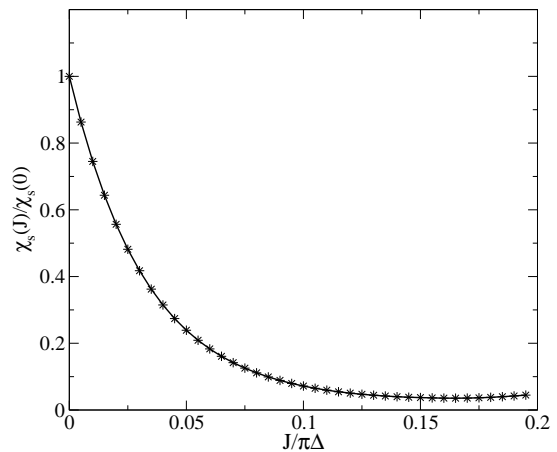}
     \caption{(Color online) A plot of $ \chi_s(J)/ \chi_s(0)$
as a function of
 $J/\pi\tilde\Delta$ for $U_{12}=U+3J/2$, $U/\pi\Delta=5$.}
     \label{chis_hund}
   \end{center}
 \end{figure}

In the antiferromagnetic regime there is still
the tendency for  the impurity spins to be screened more locally as $J$
is increased. This can be seen in the plot of $\chi_s$ shown in
 Fig. \ref{chis_hund}. The value of $\chi_s$ decreases monotonically as
 $J/\pi\Delta$ increases to a value  $J/\pi\Delta\sim 0.15$.
After that point there appears to be a slight upturn, but no real significance
can be attached to this as it involves the product of a quantity tending
to zero $\tilde\eta_s$ and a diverging quantity $1/z\pi\Delta$ as $z\to 0$,
and numerical errors start becoming significant when $z$ becomes very small.
Though we get no transition in this case it does not correspond to the
category described by Affleck et al. \cite{ALJ95}, where the phase shift
$\delta$ decreases with increases of $J$. For the whole range of $J$ the phase
 shift remains at the value $\delta=\pi/2$. It would appear that provided the
Friedel sum rule applies for particle-hole symmetry $\langle
 n_{d,\alpha,\sigma}\rangle =1/2$ and if Eq. $\epsilon_d+\Sigma(0)=0$, 
 we always have a phase shift  $\delta=\pi/2$, so we cannot have a continuous
 crossover from  $\delta=\pi/2$ to
 $\delta=0$. Either there is no transition and $\delta=\pi/2$ for all $J$ or
 the
self-energy develops a singularity at $\omega =0$, with a breakdown of the
Friedel sum rule, and a sharp transition to a state with  $\delta=0$.
\par

 \noindent
 \noindent

 \noindent
 \noindent

\section{Conclusions}

We have examined the low energy behavior in the Fermi liquid regime of a model
that has two types of quantum critical points. The thermodynamic and dynamic
response functions  in this regime can be expressed  in terms of a limited
number of renormalized parameters, $\tilde\Delta$, $\tilde U$, $\tilde J$ and
$\tilde U_{12}$, which we can calculate accurately from an analysis of the NRG
low energy fixed point. Once these have been determined they can then be
substituted into the relevant RPT formulae. The fact that certain
susceptibilities
remain finite at the transition, where the quasiparticle weight $z\to 0$,
gives enough equations to predict the dimensionless parameters, $\tilde U\tilde\rho(0)$,
 $\tilde J\tilde\rho(0)$ and  $\tilde U_{12}\tilde\rho(0)$ in the Fermi liquid
 region on the approach to the critical point. These predictions have been
 confirmed
from the NRG results for both types of transition, including situations away
from particle-hole symmetry, and dot asymmetry in the case of the transition
at $J=J_c$. As these dimensionless parameters are
universal on the approach to the critical points, the quasiparticle interactions,  $\tilde U$,
 $\tilde J$ and  $\tilde U_{12}$, can all be expressed in terms of a single
 energy scale $T^*=1/4\tilde\rho(0)$, where $T^*\to 0$ at the critical point. \par
In this paper we have not examined the finite temperature non-Fermi liquid
regime in the region of the critical point at $J=J_c$, but in our NRG results
we found the same higher energy non-Fermi liquid fixed point as observed in
previous calculations for the two impurity Kondo model\cite{ALJ95}. Those results were
explained by Affleck and Ludwig\cite{AL92,ALJ95} using conformal field theory. More recent
work has clarified with the relation between this transition and the non-Fermi
liquid fixed point of the two channel Kondo model. A recent paper of 
Sela and Mitchell\cite{MS12} has addressed the crossover as a function of temperature
between the Fermi liquid and non-Fermi liquid regimes. It would be interesting
to examine this crossover in terms of renormalized parameters using the
approach which was  developed and applied to the two channel Kondo model\cite{BHZ97,BBHZ99}.\par

It has proved difficult to probe the quantum critical point of the two
dot/impurity  model experimentally. The problem is that in any experimental
set up there will be in addition to the interdot/impurity interaction a direct
or indirect hybridization term. This was the case in the recent experimental
work on a two impurity system which consisted of a cobalt ion on a STM point
and interacting with a second cobalt ion  on a
metal surface. The hybridization term destroys the critical
point. Nevertheless it would be interesting to generalize the model used here
to model this experimental system, to see if the quantum critical behavior
could be observed due to the presence of a higher energy non-Fermi liquid
fixed point, as was done in the case of the two channel Kondo model.\par
The quantum critical points studied here are for impurity models and cannot
be applied directly to the experimental results on quantum critical
behavior observed in lattice heavy fermion systems. However, certain 
features found may be universal and apply to a general class quantum critical 
points. For example, the fact that on the approach to the quantum critical
points all the quasiparticle interactions could be expressed in terms of a
single energy scale $T^*$ means that any dynamic response function
$\chi(\omega)$ should take the form $\chi(\omega)=F(\omega/T^*,T/T^*)$.
At the critical point where $T^*\to 0$ we would then expect a form, 
$\chi(\omega)=T^af(\omega/T)$, or $\omega/T$ scaling which has been observed
at the several heavy fermion  critical points, such as in YbRu$_2$Si$_2$\cite{Pfau2012}. 
The quantum critical point
induced in  YbRu$_2$Si$_2$ by suppressing the antiferromagnetic order with a magnetic
field and has been interpreted as a Kondo
breakdown\cite{SS10,CPSR01},  where the loss of the f-electron states at the Fermi surface
results in a change in volume of the Fermi surface at the quantum critical
point. A schematic sketch is given Fig. 4 in the paper of Pfau et
al. \cite{Pfau2012} of the evolution of the quasiparticle weight factor $z$
across such a quantum critical point.  At the critical point $J=J_c$ in our
calculations in the particle-hole symmetric case  there is a similar Kondo breakdown because the Kondo resonance
at the Fermi level for $J<J_c$ collapses at $J=J_c$ and there are no
impurity/dot 
f-states at the Fermi level for $J>J_c$.
In  Fig. \ref{delratio1} we have  precise
results for this evolution of the quasiparticle weight factor $z$ for
the Kondo collapse in a two impurity model (note that for comparison with the
sketch for YbRu$_2$Si$_2$,
 the range
$J>J_c$ corresponds to the small Fermi surface and $J<J_c$ to the large one).\par

\bigskip
\noindent{\bf Acknowledgment}\par
\bigskip
{We thank Akira Oguri and Johannes Bauer for helpful discussions. Two of us
  (DJGC and ACH) thank the EPRSC for support (Grant No. EP/G032181/1).
The numerical calculations were partly carried out at the Yukawa Institute Computer Facility.
\par
\bigskip

\section{Appendix}

We clarify here  how the analysis of the low  energy NRG fixed points which we
use in this paper differs from the approaches based on the  work of
Krishnamurthy, Wilkins and Wilson (KWW I and II) \cite{KWW80a,KWW80b}.  We consider
the case of the single impurity Anderson model.  For the particle-hole
symmetric
Anderson model in the KWW approach the low energy NRG fixed
point is taken to correspond to the free conduction chain with the impurity
and first conduction site removed. This  corresponds to $V\to\infty$ in the
bare model, and the equivalent of the strong
coupling fixed point taken for the Kondo model in the original calculation
of Wilson \cite{Wil75}, as the Kondo coupling $J\sim V^2/U$.   
 The low energy fixed point, however, can equally well
be viewed as the fixed point of a non-interacting Anderson model with
a finite hybridization $\tilde V\ne 0$, which we shall denote as the Fermi liquid (FL)
fixed point.
 The fixed point corresponding to a free NRG conduction chain
 depends only on whether the chain has an even or odd number of sites. As two
sites are removed from the original NRG chain in interpreting the fixed point
in the KWW approach, the KWW and FL fixed points are equivalent.\par

When the leading irrelevant corrections to the fixed point are taken into
account in the KWW calculation, the effective Hamiltonian takes the form,
\begin{equation}H_{N,eff}=H_{N,SC}+\omega_1\Lambda^{(N-1)/2}\delta
  H_1+2\omega_2\Lambda^{(N-1)/2}\delta H_2, \end{equation}
with 
\begin{equation}\delta H_1=\sum_\sigma
(c^{\dagger}_{1,\sigma}c^{}_{2,\sigma}+c^{\dagger}_{1,\sigma}c^{}_{2,\sigma}),\quad
\delta H_2=(\sum_\sigma n_{1,\sigma}-1)^2, \end{equation}
where $H_{N,SC}$ is the Hamiltonian for the truncated free conduction chain,
$ \delta H_1$
corresponds to a  correction to the hopping matrix
element between the first and second conduction sites of the truncated
conduction chain,
and $\delta H_2$, an
on-site interaction at the first site of the truncated chain. The corrections
to the free energy are then calculated to first order in $\omega_1$ and
$\omega_2$, and the following expression derived for the impurity specific heat
coefficient, 
\begin{equation}
\gamma=-{\omega_1\over D}\left({4\over
    1+\Lambda^{-1}}\right){\alpha_0\alpha_1\over {\rm
    ln}\Lambda},
\label{wilgam}\end{equation}
in units of $2\pi^2k_{\rm B}^2/3$, 
and for the susceptibility,
\begin{equation}
\chi=-{\omega_1\over D}\left({2\over
    1+\Lambda^{-1}}\right){\alpha_0\alpha_1\over {\rm
    ln}\Lambda}+{\omega_2\over D}\left({2\over
    1+\Lambda^{-1}}\right){\alpha^4_0\over ({\rm
    ln}\Lambda)^2}
\label{wilchi}
\end{equation}
in units of $(g\mu_{\rm B})^2$, 
 where
\begin{equation}\alpha_0^2={1\over 2}(1-\Lambda^{-1}), \quad \alpha_1^2={1\over 2}(1-\Lambda^{-3}). \end{equation}

 The values of $\omega_1$, $\omega_2$ in the KWW approach have to be
 calculated from the asymptotic approach of the energy levels to their
fixed point values.  If $E_p(N)$
is the energy of the lowest single-particle excitation from the ground state
in the NRG calculation for a chain length $N$, and
$E^*_p$ is the corresponding value at the fixed point, 
the value of $\omega_1$
 can be obtained by matching the difference
 $E_p(N)-E^*_p$ to that obtained from fixed point Hamiltonian with the leading
 order correction term $\delta H_1$ for large $N$.  
Similarly $\omega_2$  can be calculated from the difference
between the lowest two-particle excitation $E_{pp}(N)$ and the lowest two single-particle
excitations  $E_p(N)$ for large $N$. \par

We contrast this with the FL analysis where the two leading order corrections correspond to the
effective hybridization with the impurity $\tilde V$ and the effective
on-site interaction  $\tilde U$ at the impurity site. The hybridization term, 
though a leading irrelevant term, it is technically a {\em dangerously
  irrelevant }
term, because the fixed point changes when  $\tilde V=0$  from that corresponding to an even (odd) chain to that
for an odd (even) one. 
The value of $\tilde V$ can be obtained by requiring  $E_p(N)$
for large $N$  correspond to the lowest single particle excitation of a
{\em non-interacting} Anderson model with hybridization parameter $\tilde V$,
as described briefly in section IV.  The corresponding value of $\tilde \Delta$ for the  continuum model
  is given by $\tilde\Delta=\pi\tilde V^2/D A_{\Lambda}$, where
\begin{equation}
 A_\Lambda={1\over 2} {\rm ln}\Lambda \left[{ 1+\Lambda^{-1}\over
  1-\Lambda^{-1}}\right]
\label{Alam}\end{equation}
is a correction factor to the bandwidth $D$ on using a discretized NRG
chain with discretization
parameter $\Lambda (>1)$.  
Once $\tilde V$ has been determined $\tilde U$ (and $\tilde J$) can be
calculated from the difference between the lowest two-particle and the lowest
two single-particle excitations in a similar way to the calculation of
$\omega_2$. 
Once the renormalized parameters have been determined the physical quantities,
such as the specific heat coefficient $\gamma$, and the spin and charge
susceptibilities, can be calculated by substituting into the relevant RPT
formulae. For further details we refer to references \cite{HOM04,NCH10s}.\par

We can establish a connection between the two approaches using the  equations
given in Appendix E of the  KWW I
\cite{KWW80a}.
In this Appendix $\omega_1$ and $\omega_2$ are calculated to first order
in $U$ for the symmetric Anderson model.  In the RPT the exact results
for $\gamma$ and $\chi$ correspond to such a calculation but with
{\em renormalized}  parameters so we can use the results given in equations (5.38)
and (5.39) in KWW I by replacing
$\Delta$ (or $\Gamma$ in the notation used in the paper) by
$\tilde\Delta$, and $U$ by $\tilde U$, which gives
\begin{equation}\omega_1=-{1\over 2}[{1\over 2}(1+\Lambda^{-1})]^2{({\rm ln}\Lambda)^2
\over \alpha_0^3\alpha_1}{D\over 2\pi\tilde\Delta},\end{equation}
\begin{equation}
\omega_2= [{1\over 2}(1+\Lambda^{-1})]^3 {({\rm ln}\Lambda)^4
\over \alpha_0^8}\left[{D\over 2\pi\tilde\Delta}\right]^2\left[{\tilde U\over
  2D}\right]
\end{equation}
We see that both $\omega_1$ and $\omega_2$ diverge as $\tilde\Delta\to 0$,
if $\tilde U/\pi\tilde\Delta$ remains finite, which explains why the
renormalized parameters we calculate on the approach to the QCP tend to zero,
and the corresponding terms on using the KWW  approach diverge.\par
 On using Eqns (\ref{wilgam}) to (\ref{Alam}) we find for the continuum model,
\begin{equation}
\gamma={2\pi^2k_{\rm B}^2\over
  3 \pi\tilde\Delta},\quad
\chi={(g\mu_{\rm B})^2\over 2\pi\tilde\Delta}\left[1+{\tilde U\over
   \pi\tilde\Delta}\right],\label{rpt_eqn}\end{equation}
which is the RPT result for the
   symmetric model.\par
There is a clear advantage in the Fermi liquid analysis for
   the asymmetric Anderson model as one  deals with the same low energy fixed
   point with just the additional renormalized parameter
   $\tilde\epsilon_{d}$, so that results in Eqn. (\ref{rpt_eqn}) are
  generalized by replacing $1/\pi\tilde\Delta$ by $\tilde\rho(0)$.
 The corresponding KWW analysis requires consideration
   of a number of low energy fixed points, strong coupling, intermediate
   valence, etc, to cover the full parameter range. The generalized results
for Eqn. (\ref{rpt_eqn}), however, can be deduced 
from Appendix D in KWW II\cite{KWW80b} with the substitutions, $U\to \tilde U$,
 $\Gamma\to \tilde\Delta$ and $\epsilon_d+U/2\to \tilde\epsilon_d$
in Eqns (D.23-25).

\par

\bibliography{artikel}
\bibliographystyle{h-physrev3}

\end{document}